\documentclass[aps,pre,final,letterpaper,10pt,twocolumn,floats,showpacs,amsmath,amsfonts,amssymb]{revtex4-1}
\usepackage{graphicx} % ZAWSZE PRZED epstopdf
\usepackage{epsf}
\usepackage{epstopdf}
\usepackage{tikz}
\usepackage{float}
\usepackage[colorlinks=true,citecolor=blue]{hyperref}
\usepackage[normalem]{ulem}

\begin{document}

\title{Statistical correlations of currents flowing \textit{via} proximized quantum dot}

\author{G. Micha{\l}ek} \email{grzechal@ifmpan.poznan.pl}
\author{B. R. Bu{\l}ka}
\affiliation{Institute of Molecular Physics, Polish Academy of Sciences, ul. M. Smoluchowskiego 17, 60-179 Pozna\'{n}, Poland}

\author{T. Doma\'{n}ski}
\author{K. I. Wysoki\'{n}ski}
\affiliation{Institute of Physics, M. Curie-Sk{\l}odowska University, pl. M. Curie-Sk{\l}odowskiej 1, 20-031 Lublin, Poland}

\date{\today}

\begin{abstract}
Statistical properties of the electron transport flowing through nanostructures are strongly influenced by the interactions, geometry of the system and/or by type of the external electrodes. These factors affect not only the average current induced in the system but also contribute to fluctuations in the flux of charges and their correlations. Due to possible applications of the hybrid nano-systems, containing one or more superconducting electrodes, a detailed understanding of the flow of charge and its fluctuations seems to be of primary importance. Coulomb repulsion between electrons usually strongly affect the current-current correlation function. In this work we study the correlations in the charge flow through such interacting quantum dot contacted to one superconducting and two normal electrodes. This set-up allows for analysis of the Andreev scattering events in the correlations of currents flowing between external electrodes and, in particular, gives access to cross-correlations between currents from/to different normal electrodes.
Our approach relies on the master equation technique, which properly captures the Coulomb interactions.
We study the finite frequency correlations and find the relaxation processes, related to the high frequency charge and low frequency polarisation fluctuations. The multiterminal structure of here studied single electron device allows to analyze a competition between the intra- and inter-channel correlations. In the appropriate limit of the interacting quantum dot embedded between two normal electrodes our calculations quantitatively describe the recent experimental data on the frequency dependent correlations. This shows a promising potential of the method for description of the hybrid systems with superconducting electrode(s).
\end{abstract}

% 74.45.+c Proximity effects; Andreev reflection; SN and SNS junctions
% 73.23.Hk Coulomb blockade; single-electron tunneling
% 72.70.+m Noise processes and phenomena
% 73.63.Kv Quantum dots

\maketitle

\section{\label{sec:intro} Introduction}

Studies of charge transport through the multi-terminal nanostructures with quantum dots are important from both, practical and fundamental science points of view. Such systems have been proposed as \textit{e.g.} efficient heat to electricity converters~\cite{benenti2017} and sources of entangled electrons~\cite{franceschi2010} in heterojunctions made of the normal and superconducting electrodes. Ability to control their microscopic parameters, such as the energy levels of quantum dots, interactions between opposite spin electrons and the coupling of quantum dot to external leads makes such heterostructures very appealing. Furthermore, the character of transport can be changed in such nanostructures from the sequential to ballistic tunneling upon varying the macroscopic parameters, \textit{e.g.} dot size, resistance of the tunnel junctions, gate voltage and bias, temperature, magnetic field, \textit{etc}.~\cite{wiel2003, zumbuhl2004}. They are thus important playground for studying many-body phenomena, such as the Kondo effect, competition between the local and non-local Andreev reflections and many other~\cite{franceschi2010, mazza2015}. Some of these phenomena might find potential applications in nanoelectronics and spintronics~\cite{braunecker2013} or become basic building blocks of future quantum computers~\cite{zoller2005, rodero2011, eschrig2011}.

Tunneling processes through a quantum dot embedded between one superconducting and several normal electrodes can be contributed by a number of different events. Electrons may tunnel between the normal electrodes -- we call this process electron transfer (ET). Another possible processes rely either on the direct (D) and crossed (C) Andreev reflections (AR), that are the main subject of our study here. Of special interest for applications are the latter processes (CAR), in which two electrons of the strongly entangled BCS singlet state are scattered into different normal electrodes. Such events have been proposed as a source of the spatially separated entangled electrons for potential use in quantum computing~\cite{rodero2011}. High efficiency of the Cooper pair splitting can be achieved in the systems with strongly interacting quantum dots and detected by measuring non-local differential conductance of the Andreev processes~\cite{franceschi2010, recher2001}. Correlations in the electrical currents have indeed provided evidence for such entanglement~\cite{russo2005, hofstetter2009, schindele2012, braunecker2013}.

Besides valuable analysis of the conductances one can get additional information about properties of the system, concerning \textit{e.g.} mechanism of transport, statistics of quasiparticles contributing to charge transport, role of the Coulomb interactions, relaxation processes, correlations between currents flowing \textit{via} different transport channels and/or different branches of the device by studying the shot noise, \textit{i.e.} time-dependent fluctuations in electrical currents caused by the charge quantization~\cite{blanter2000}. In the systems with noninteracting electrons the Pauli exclusion principle is known to cause the antibunching~\cite{feynman1965}, that manifests itself by reduction of the shot noise (autocorrelations) below the Poissonian value $S_P = 2 e J$ (where $J$ is an average current flowing through the system) or by negative cross-correlations in the multi-terminal systems~\cite{blanter2000}. It has been found, however, that correlations between the tunneling electrons can also lead to bunching as evidenced by the super-Poissonian noise~\cite{iannacone1998, bulka1999, bulka2000}. In the multi-terminal structure the dynamical channel blockade could be responsible for enhancement of the shot-noise~\cite{michalek2002, bulka2008} and for the positive cross-correlations~\cite{michalek2009, bulka2008, dong2009}. Such bunching and anti-bunching features have been indeed observed experimentally (in the auto- and cross-correlations) by McClure \textit{et al.}~\cite{mcclure2007} and Zhang \textit{et al.}~\cite{zhang2007} for a device, comprising the double quantum dots that are coupled capacitively. Recent measurements \cite{ubbelohde2012} of the correlations of currents flowing through the interacting quantum dot contacted by two normal electrodes have shown the ability of experiments to investigate the fingerprints of interactions and the coherence, being the main theme of our work. We shall comment on this experiment in Sec.~\ref{sec:freq}. However, our system containing the superconducting electrode allows the study of not only the cross-correlations but also the superconducting coherence of electrons on the frequency dependent Fano factors and other characteristics.

Dynamical correlations have been also studied in electrical currents of superconducting systems, including chaotic cavities~\cite{borlin2002, morten2008}, planar junctions with direct normal metal--superconductor interfaces~\cite{anantram1996,torres1999,bignon2004,freyn2010,wei2010,floser2013,maisi2014,albert2016,gil2017,golubev2019}, topological wires~\cite{valentini2016,komnik2016} and quantum dot systems in two-~\cite{zhao2002,gogolin2006,zhang2009,soller2011,braggio2011,droste2015,dong2017,pistolesi2015} or three-terminal (Cooper pair splitter)~\cite{das2012, weiss2017, trocha2018, walldorf2018} configurations, \textit{etc}. In particular, it has been shown that positive cross-correlations in hybrid systems are a signature of the high efficiency of Cooper pair splitting~\cite{das2012}.
Most of the calculations have so far explored the zero-frequency limit~\cite{zhao2002, gogolin2006, soller2011, braggio2011, dong2017, weiss2017, trocha2018} in two-terminal setups~\cite{zhao2002, gogolin2006, zhang2009, soller2011, braggio2011, droste2015, dong2017} neglecting the Coulomb interactions~\cite{zhao2002, zhang2009}. The short-time dynamics have been recently addressed by means of factorial cumulants in a metallic single-electron box~\cite{stegmann2016} or employing the waiting time distribution approach to the case of unidirectional transport, \textit{i.e.} for very large biases~\cite{rajabi2013, walldorf2018, michalek2018}.

In this paper we investigate the noise of three-terminal hybrid system with a quantum dot embedded in Y-shape configuration, between one superconducting and two metallic leads. For this nanostructure we analyze an interplay between the tunneling of normal electrons and the Andreev reflection processes evidenced in the auto- and cross-correlations between tunneling currents and in the corresponding Fano factors. Our considerations go beyond the zero-frequency noise, capturing also finite-frequency contributions due to charge fluctuations between the QD and normal tunnel junctions. We restrict here to the subgap regime (neglecting relaxation processes by the quasiparticles from outside the pairing gap of superconducting lead), therefore this study is practically valid up to milielectronvolt (infrared frequencies) region. Furthermore, we do not address any short-time coherent oscillations of the electron pairs between the QD and superconducting reservoir~\cite{baumgartner2017}. The zero-frequency noise explored previously in the literature describes total fluctuations, which are rather noisy. Our studies extended onto the finite-frequency range give an insight into the dynamics of electron and hole tunneling processes and their correlations. We are thus able to single out from such noisy spectrum the relaxation processes contributed by the individual subgap quasiparticles.

Our study is based on the master equation method, reliable for the incoherent tunneling regime  $k_{B}T \gg \Gamma_{L(R)}$ (where $\Gamma_{L(R)}$ denotes tunneling rate between QD and the metallic lead).
This analysis of frequency-dependent noises for a three-terminal hybrid device with the proximized QD significantly extends earlier studies of two-terminal systems, using the diagrammatic real-time approach and the generalized master equation~\cite{droste2015, braggio2011}. To facilitate some comparison with former studies and to emphasize the role played by second normal electrode we show also the numerical results obtained for 2-terminal N-QD-S system.

The paper is organized as follows. In Sec.~\ref{sec:model} we introduce the microscopic model, describing the QD strongly coupled to superconducting reservoir and weakly coupled to two metallic electrodes. Next, in Sec.~\ref{sec:method},
% we discuss the electron and hole currents  and
determine the frequency-dependent current-current correlation functions of the electron and hole charge transport through the in-gap (Andreev) bound states. In Sec.~\ref{sec:results} we present the numerical results, considering the case of small and large biases. We analyze in detail the frequency dependent Fano factors and current correlations, monitoring contributions from the currents through various Andreev bound states (ABS), what gives insight into internal dynamics of the system. In Sec.~\ref{sec:freq} we discuss the frequency dependent Fano factors relevant to our model and present their comparison to available experimental data \cite{ubbelohde2012} on the quantum dot coupled between two normal terminals. In Sec.~\ref{sec:summary} we summarize the main findings and, in Sec.~\ref{sec:concl-outl}, outline some future perspectives related to our study.

\section{Model and methodology}

\subsection{\label{sec:model} Microscopic model}

\begin{figure}[thb] % Fig 1
\includegraphics[width=0.85\linewidth,clip]{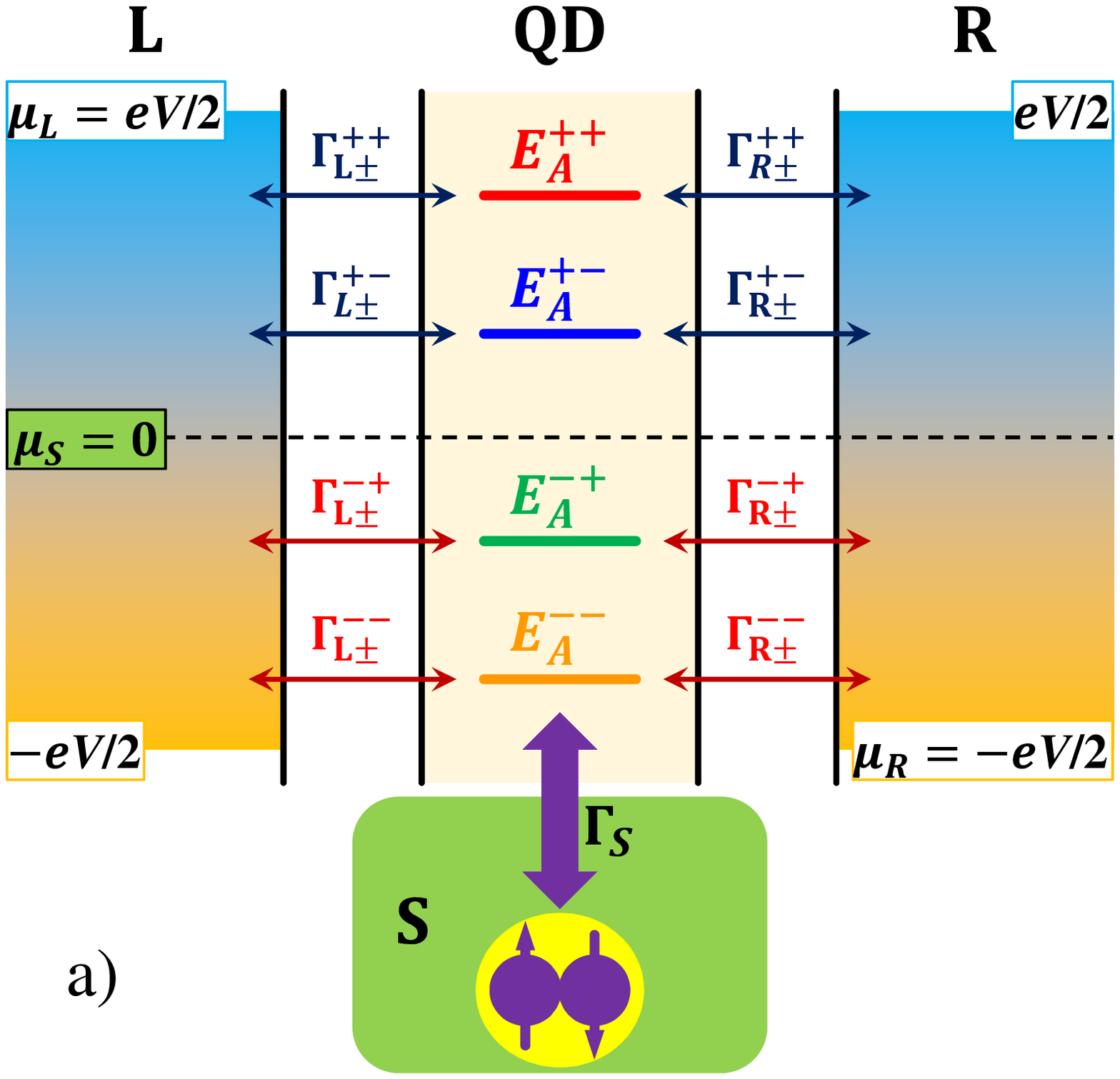} \\
\includegraphics[width=0.85\linewidth,clip]{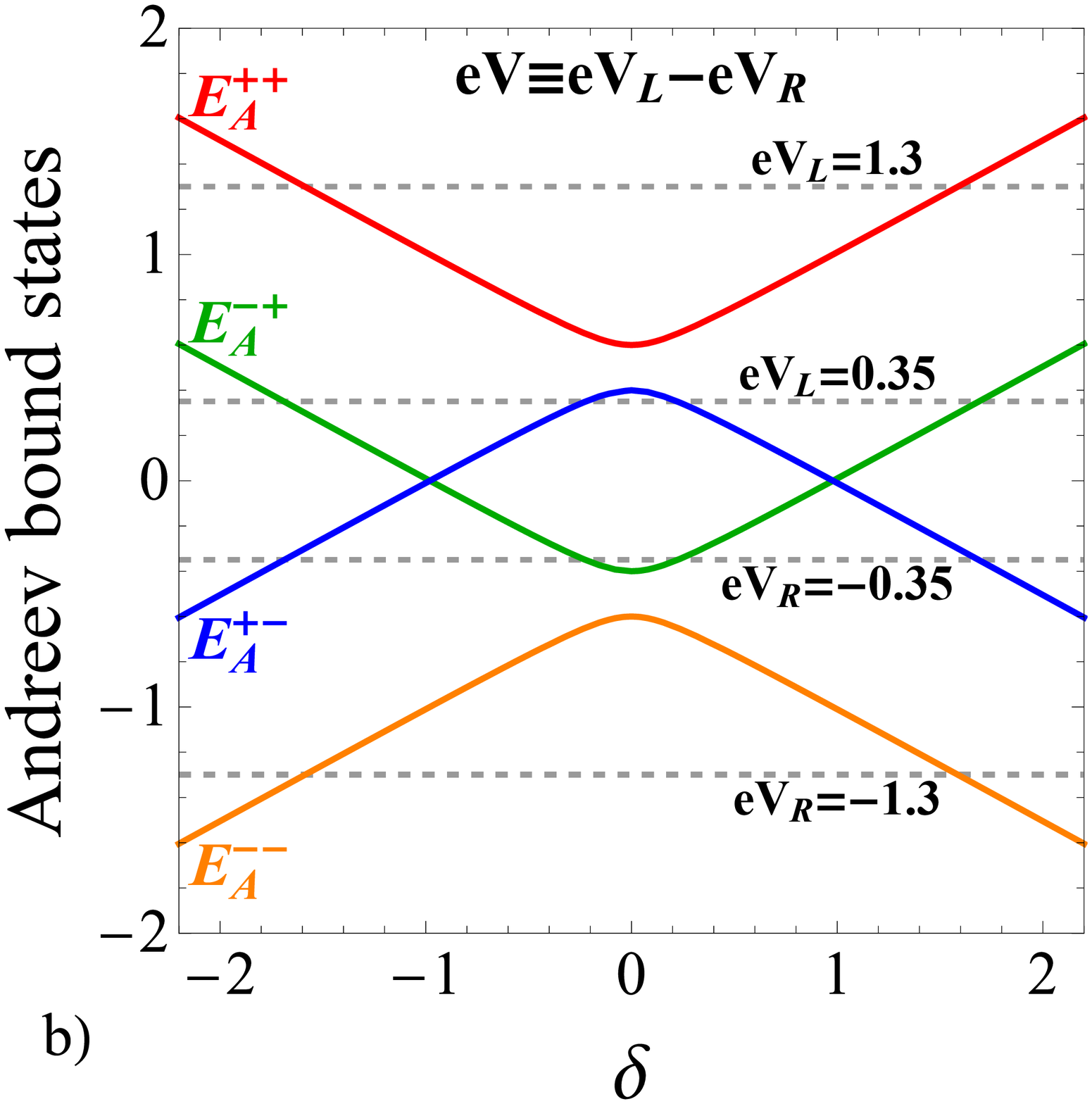}
\caption{(a) Schematic view of the hybrid structure, comprising the proximized quantum dot strongly coupled to the superconducting (S) electrode and weakly coupled to the metallic left (L) and right (R) normal leads.
Tunneling rates through the Andreev bound states $E_A^i$ ($i \in \{ ABS \} \equiv \{ ++, -+, +-, -- \}$) related to quasiparticle excitations between the  even$\leftrightarrow$odd eigenstates are indicated by the arrows. For energies above the chemical potential of  superconductor ($\mu_S=0$) the charge carriers are mainly electrons, while below it the holes are dominant. (b) Dependence of ABS on the detuning $\delta = 2 \epsilon_1 + U$  from the half-filled QD. The horizontal dashed lines denote a small ($eV \equiv eV_L - eV_R = 0.7$) and large ($eV = 2.6$) bias windows obtained in Sec. \ref{sec:results} for $\Gamma_S = 0.2$, treating the Coulomb potential as the energy unit $U\equiv 1$.}
\label{fig1}
\end{figure}

We consider a single level quantum dot (QD) strongly hybridized with a superconducting lead (S) and weakly coupled between the left (L) and right (R) normal metallic electrodes, as depicted in Fig.~\ref{fig1}(a). For simplicity we shall focus on the \textit{superconducting atomic limit}, assuming that a pairing gap $\Delta$ of the superconducting reservoir is the largest energy scale in our study. This assumption allows us to neglect any single particle tunneling to/from the superconductor, restricting solely to the subgap tunneling processes. Under such circumstances, an effective Hamiltonian describing the proximized QD takes the BCS-type form~\cite{vecino2003}
\begin{equation}
\label{eq:heff}
H_{eff} = \epsilon_1 \sum_{\sigma} d_\sigma^\dag d_\sigma + U n_\uparrow n_\downarrow - \frac{\Gamma_S}{2} \left( d_\uparrow^\dag d_\downarrow^\dag + d_\downarrow d_\uparrow \right) \; ,
\end{equation}
where $\epsilon_1$ is the spin-degenerate energy level, $d_\sigma^\dag (d_\sigma)$ create (annihilate) electron with spin $\sigma = \{ \uparrow, \downarrow \}$, $n_\sigma \equiv d_\sigma^\dag d_\sigma$ stands for the number operator, and $U$ denotes the repulsive Coulomb interaction. The last term in Eq.\ (\ref{eq:heff}) describes the induced pairing between the opposite spin electrons originating from the Cooper pairs leaking onto the QD (superconducting proximity effect). Efficiency of such process is in the superconducting atomic limit controlled by the coupling $\Gamma_S$~\cite{weiss2017}.

The proximized QD, Eq.~(\ref{eq:heff}), mixes the empty $| 0 \rangle$ with doubly occupied $| D \rangle \equiv | { \uparrow \downarrow } \rangle$ configurations. True eigenstates are thus represented by the coherent superpositions $| - \rangle = \alpha_{+} | 0 \rangle + \alpha_{-} | D \rangle$ and $| + \rangle = \alpha_{-} | 0 \rangle - \alpha_{+} | D \rangle$ with the eigenenergies $\epsilon_{\pm} = \delta / 2 \pm \epsilon_A$, where the BCS coefficients are  $2\alpha_{\pm} = \sqrt{2 \pm \delta /\epsilon_A}$. For convenience, we have introduced $2 \epsilon_A = \sqrt{\delta^2 + \Gamma_S^2}$ accounting for the energy splitting between $| + \rangle$ and $| - \rangle$ states, whereas $\delta = 2 \epsilon_1 + U$ describes detuning between $| 0 \rangle$ and $| D \rangle$ states~\cite{wysokinski2012,rajabi2013,braggio2011,droste2015}.
Besides these even states $|\pm\rangle$ there exists also a subspace of the odd (singly occupied) states $| { \uparrow } \rangle$ and $| { \downarrow } \rangle$ with the degenerate eigenenergy $\epsilon_1$ (unless a magnetic field is applied).

Charge transport between the normal electrodes via the proximitized QD is strictly related to quasiparticle transitions between the even and odd eigenstates. Optimal conditions for the subgap conductance occur when the bias voltage $V$ is tuned to energy difference between the initial and final state. Such quasiparticle excitation energies (depicted in Fig.~\ref{fig1}) define a set of the Andreev bound states~\cite{hewson2007} $E_A^{++} \equiv \epsilon_+ - \epsilon_1 = \epsilon_A + U / 2$, $E_A^{+-} \equiv \epsilon_- - \epsilon_1 = - \epsilon_A + U / 2$, $E_A^{-+} \equiv \epsilon_1 - \epsilon_- = \epsilon_A - U / 2$, and $E_A^{--} \equiv \epsilon_1 - \epsilon_+ = -\epsilon_A - U / 2$. The gate voltage or $\delta$ dependence of the Andreev states are shown in Fig.~\ref{fig1}(b). Depending on the source-drain voltage $eV = \mu_L - \mu_R$, also marked in the figure, two or more Andreev bound states participate in the transport. The chemical potential of the left (right) normal electrode is denoted by $\mu_L$ ($\mu_R$). We shall consider two situations, corresponding to small ($eV = 0.7$) and large ($eV = 2.6$) bias, respectively.

The metallic leads are represented by the free fermions
\begin{equation}
\label{eq:hleads}
H_\alpha = \sum_{k, \sigma} (\epsilon_{\alpha k}-\mu_\alpha) c_{\alpha k \sigma}^\dag c_{\alpha k \sigma} \; ,
\end{equation}
where $c_{\alpha k \sigma}^\dag (c_{\alpha k \sigma})$ creates (annihilates) itinerant electron with spin $\sigma = \{ \uparrow, \downarrow \}$ and momentum $k$ in the lead $\alpha = \{ L, R \}$. Hybridization of the proximized QD, Eq.~(\ref{eq:heff}), with both external metallic electrodes is given by
\begin{equation}
\label{eq:htun}
H_T = \sum_{\alpha, k, \sigma} \left( t_\alpha c_{\alpha k \sigma}^\dag d_\sigma + t_\alpha^* d_\sigma^\dag c_{\alpha k \sigma} \right) \; .
\end{equation}
Since we are interested in the low energy physics, safely smaller than the superconducting energy gap $\Delta$, it is convenient to introduce the tunneling rates $\Gamma_\alpha$ describing electron and hole transfer between the QD and metallic leads. In the wide-band limit approximation these tunneling rates $\Gamma_\alpha = 2 \pi \sum_k |t_\alpha|^2 \delta ( E - \epsilon_{\alpha k} ) = 2 \pi |t_\alpha|^2 \rho_\alpha$, where $\rho_\alpha$ is the density of states in the normal metal lead $\alpha = \{ L, R \}$, can be approximated by the constant parameters.

\subsection{\label{sec:method} Currents and noise}

We are interested in dynamical fluctuations of the current from its averaged value, $\Delta \hat{J}_{\alpha}(t)\equiv \hat{J}_{\alpha}(t) -\langle J_\alpha(t) \rangle $. Definitions of these currents are presented in Appendix~\ref{appA}. To describe the current fluctuations in the contact $\alpha$ and $\beta$ we use the time correlation function~\cite{blanter2000}
\begin{equation}\label{eq:noiset}
S_{\alpha \beta}(t,t') \equiv \frac{1}{2} \langle \Delta \hat{J}_{\alpha}(t) \Delta \hat{J}_{\beta}(t') + \Delta \hat{J}_{\beta}(t') \Delta \hat{J}_{\alpha}(t) \rangle \; .
\end{equation}
For the case with time-independent external fields the correlation function is a function of $\tau=t'-t$, and its Fourier transform can be expressed as
\begin{equation}\label{eq:noiseomega}
S_{\alpha \beta}(\omega) = 2 \int_{-\infty}^{\infty} d\tau e^{i \omega \tau} S_{\alpha \beta}(\tau)\;,
\end{equation}
which is dubbed \textit{noise power}.

To determine the current-current correlation functions for the considered hybrid system we use the generation-recombination approach~\cite{vliet1965} and the method developed for spinless electron noise in a single electron transistor~\cite{korotkov1994}, extending it in our case to the multi-channel and multi-charge tunneling processes. According to this procedure, the correlation function between the currents flowing through $\alpha$-th and $\beta$-th junctions can be expressed as
\begin{equation}
\label{eq:sab}
S_{\alpha \beta}(\omega) = \sum_{i, j \in \{ ABS \}} S_{\alpha \beta}^{i, j}(\omega) \; .
\end{equation}
Here, we specified the contributions $S_{\alpha \beta}^{i, j}(\omega)$ to the correlation function originating from the currents $\hat{J}_\alpha^i$ and $\hat{J}_\beta^j$ through various ABS ($i, j \in \{ ++, +-, -+, -- \}$). They are formally defined by
\begin{equation}
\label{eq:sabij}
S_{\alpha \beta}^{i, j}(\omega) = \delta_{\alpha \beta} \delta_{ij} S_\alpha^{Sch, i} + S_{\alpha \beta}^{c, i, j}(\omega) \; ,
\end{equation}
where $S_\alpha^{Sch, i} = 2 e (I_{\alpha +}^i + I_{\alpha -}^i)$ is the high-frequency $\omega \rightarrow \infty$ limit of the shot noise, dubbed the Schottky noise. The Schottky term corresponds to the classical uncorrelated Poissonian transitions. The frequency-dependent part is given by
\begin{widetext}
\begin{align}
\label{eq:scabij}
S_{\alpha \beta}^{c, i, j}(\omega) = \pm 2 e^2 \sum_{m, n} & \left[ \mathop{\sum_{m' > m}}_{m'' < m} ( M_{m'm}^{\alpha, i} - M_{m''m}^{\alpha, i} ) \; G_{mn}(\omega) \mathop{\sum_{n' > n}}_{n'' < n} ( M_{nn'}^{\beta, j} p^0_{n'} - M_{nn''}^{\beta, j} p^0_{n''} ) \right. \nonumber \\
& + \left. \mathop{\sum_{m' > m}}_{m'' < m} ( M_{m'm}^{\beta, j} - M_{m''m}^{\beta, j} ) \; G_{mn}(-\omega) \mathop{\sum_{n' > n}}_{n'' < n} ( M_{nn'}^{\alpha, i} p^0_{n'} - M_{nn''}^{\alpha, i} p^0_{n''} ) \right] \; ,
\end{align}
\end{widetext}
where \textbf{M} is the matrix entering the master equation (\ref{eq:master}) describing the system studied here, $G_{mn}(\omega) = ( i \omega \mathbf{1} - \mathbf{M} )_{mn}^{-1} - p_m^0 / i \omega$ is the matrix Green's function of the proximized QD. The parameters $p^0_n$ are the stationary solutions of the master equation. We refer the reader to Appendix~\ref{appA} for a detailed discussion of the functions $M_{mn}^{\alpha, i}$ which are the off-diagonal elements of the matrix \textbf{M} related to the currents contributed through $\alpha$-th junction via $i$-th Andreev bound state. The sign ($+$) refers to the cross-correlation function between the currents from different (L and R) leads, while the opposite sign ($-$) corresponds to the auto-correlations between the currents from the same lead.

We have performed numerical calculations for the diagonalized master equation. Such approximation is valid for the strongly proximized QD ($\Gamma_S \gg \Gamma_{L (R)}$), when the subgap Andreev bound states are long-lived. We have checked that in absence of the Coulomb interactions the effective currents calculated from the diagonalized master equation (DME) are quantitatively consistent with the currents of coherent transport determined within the nonequilibrium Green function technique~\cite{michalek2013} in the limit $\Gamma_{L(R)} \ll k_BT, \Gamma_S$. In the next section we present the main results of our numerical calculations obtained for the aforementioned currents and their correlations, respectively.

\section{\label{sec:results} Results}

Deep in the superconducting gap the charge can be transmitted through our hybrid setup by one of three possible mechanisms: (i) single electron transfer (ET) between the metallic electrodes, (ii) direct Andreev reflection (DAR) when incoming electrons from the metallic lead are converted into the on-dot pairs reflecting holes back to the same normal lead, and (iii) crossed Andreev reflection (CAR), when holes are reflected to the opposite normal lead. The latter case corresponds obviously to the non-local transport processes.

In what follows we analyze the current correlations obtained for the symmetric case, when the applied voltage $V$ equally detunes the chemical potential of the left ($eV_L = eV/2$) and right ($eV_R = -eV/2$) metallic leads. The superconducting reservoir is assumed to be grounded $\mu_S = 0$. Under such circumstances contribution from the crossed Andreev reflections to the net current vanishes. Furthermore, $J_{R} = -J_{L}$ and $J_{S} \equiv 0$, therefore superconducting electrode can be regarded as a floating voltage probe~\cite{michalek2015}. Zero-frequency current fluctuations in the ballistic regime of the normal multiterminal hybrid structures (without floating superconductor) are characterized by the positive correlations~\cite{anantram1996}. It has been pointed out, however, that in the metal-superconductor-metal junctions such behavior cannot be caused by the Cooper pair splitting~\cite{floser2013,freyn2010}. In contrast, we predict here the current fluctuations in the tunneling regime, where a positive sign of the cross-correlations originates from the CAR processes~\cite{floser2013}.

For comparison with the previous studies, we also include the results obtained for the N-QD-S system (dashed lines in Figs \ref{fig2}-\ref{fig4}) determined by imposing $\Gamma_R = 0$.

\subsection{Current correlations in the small bias regime}

\begin{figure}[thb] % Fig 2
\includegraphics[width=0.86\linewidth,clip]{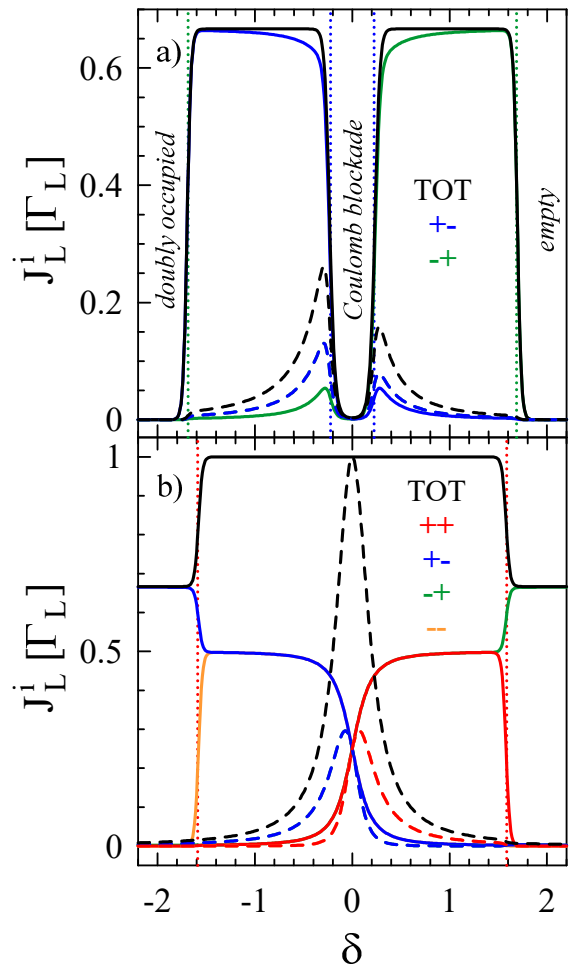}
\caption{The charge current $J_L$ (black) and its components $J_L^{++}$ (red), $J_L^{+-}$ (blue), $J_L^{-+}$ (green) and $J_L^{--}$ (orange) versus the gate voltage $\delta$ obtained for the three-terminal setup with symmetric couplings $\Gamma_R = \Gamma_L$, assuming the symmetrical bias with (a) small $eV_L = 0.35 = -eV_R$ and with (b) large $eV_L = 1.3 = - eV_R$. For comparison we also plot $J_L$ and its components for the two terminal N-QD-S case, imposing $\Gamma_R = 0$ (dashed curves). The dotted vertical lines indicate positions of the Andreev bound states $E_A^{++}$ (red), $E_A^{+-}$ (blue) and $E_A^{-+}$ (green). Numerical computations have been done for $\Gamma_S = 0.2$,  $\Gamma_R = \Gamma_L = 0.002$ and $k_BT = 0.01$.}
\label{fig2}
\end{figure}

We start, by considering the charge transport driven by a small bias $V$ which activates only two (the most inner) Andreev bound states. In Fig.~\ref{fig2} we display the total current $J_L$ and its components. For a small bias voltage [Fig.~\ref{fig2}(a)], the currents appear only for $eV_L > E_A^{+-}, E_A^{-+}$ while in the remaining regions (corresponding to the doubly occupied and empty states as well as the Coulomb blockade) the currents vanish exponentially. Furthermore, $J_L^{+-}$ and $J_L^{-+}$ are asymmetric with respect to the electron-hole symmetry ($\delta = 0$). Both currents originate predominantly from the ET processes over the entire conducting region, whereas DAR processes are enhanced only close to the Coulomb blockade (CB) region. For comparison, we have also plotted the currents obtained for the two-terminal N-QD-S nanostructure, setting $\Gamma_R = 0$ (see dashed lines), where only DAR processes are present. In this case the currents $J_L^{+-}$ and $J_L^{-+}$ are nearly identical and reveal only a small asymmetry with respect to $\delta = 0$ caused by the bias asymmetry. In the Coulomb blockade region these currents are much larger than in the three-terminal system, because in the latter case the DAR processes are suppressed by the ET tunneling.

\begin{figure*}[tbh] % Fig 3
\includegraphics[width=0.8\linewidth,clip]{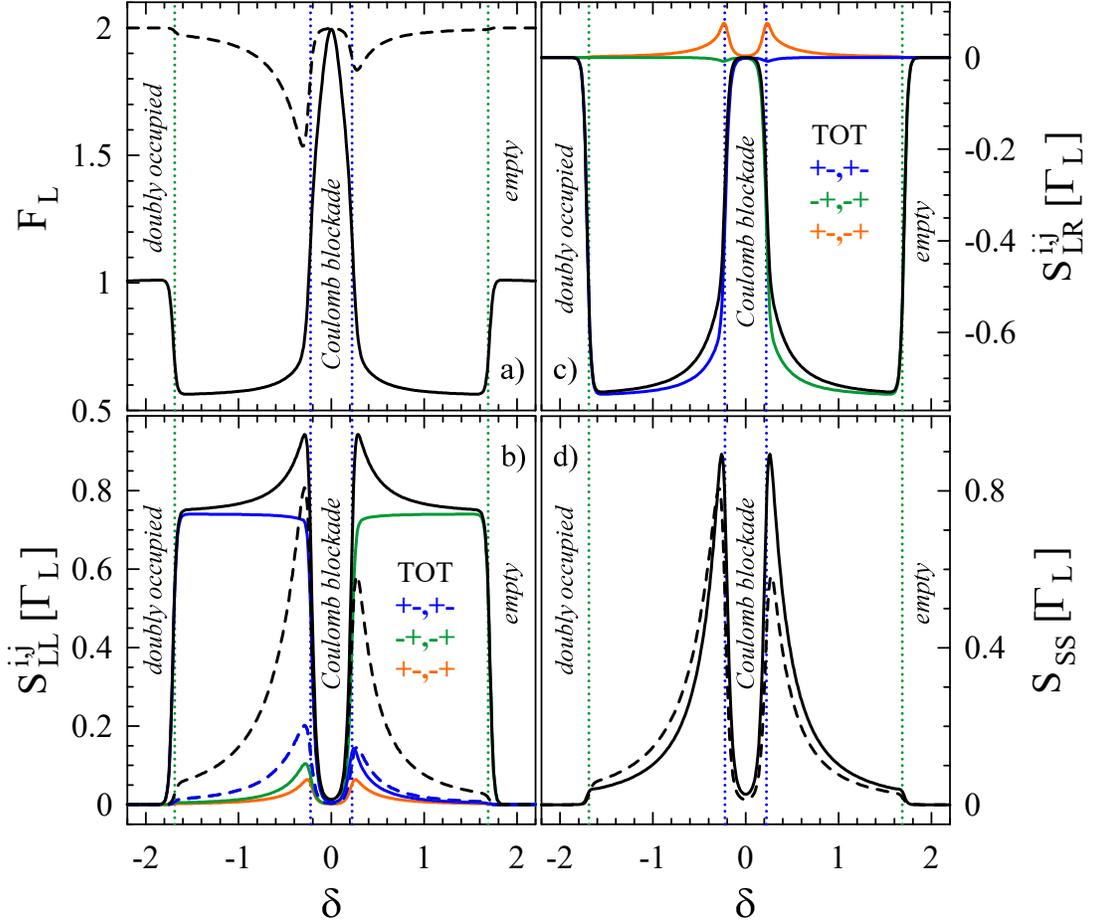}
\caption{The zero-frequency current correlations: (a) Fano factor $F_L$ for L-QD junction; (b) autocorrelation function $S_{LL}$ (black curve) and its constituents corresponding to various bound states (color curves); (c) cross-correlation function $S_{LR}$ (black curve) and its components; and (d) autocorrelation function for superconducting electrode versus the gate voltage $\delta$ obtained for small voltages $eV_L =0.35 = -eV_{R}$ assuming the symmetric couplings $\Gamma_R = \Gamma_L$. Dashed curves display results for the two terminal N-QD-S case ($\Gamma_R = 0$) and dotted vertical lines indicate positions of the Andreev bound states $E_A^{+-}$ (blue) and $E_A^{-+}$ (green). Other parameters are the same as those in Fig.~\ref{fig2}.} \label{fig3}
\end{figure*}

In Fig.~\ref{fig3} we present the zero-frequency current-current correlations. In particular, we display the zero-frequency $F_L = S_{LL}/2eJ_L$ of L-QD junction [Fig.~\ref{fig3}(a)], the autocorrelation function $S_{LL}$ [Fig.~\ref{fig3}(b)], the cross-correlation function $S_{LR}$ between different metallic leads [Fig.~\ref{fig3}(c)], and the autocorrelation function $S_{SS}$ of S-QD junction [Fig.~\ref{fig3}(d)]. We can notice, that all these quantities are symmetric with respect to $\delta = 0$. In the doubly occupied and in the empty state regime the zero-frequency Fano factor $F_L \approx 1$, indicating that such current noise is prevalent feature in the uncorrelated Poissonian tunneling of electrons. In the conducting regime the zero-frequency Fano factor diminishes $F_L < 1$ (\textit{i.e.} the noise has sub-Poissonian character), which is a signature of negative correlation of the tunneling events. For the half-filled QD ($\delta = 0$) case, the noise $F_L \approx 2$ due to uncorrelated tunneling of the Cooper pairs. The system is then in the Coulomb blockade, where single electron tunneling is suppressed at the expense of the AR processes~\cite{michalek2013, michalek2015, michalek2016}.

To get more insight into the dynamics and correlations between various tunneling processes we have also calculated the frequency-dependent noise. In the regime of unidirectional transport (\textit{i.e.} for $eV_L$ larger than $E_A^{+-}$ and $E_A^{-+}$) one can derive some analytical results. Let us underline, that in the case of small bias only two (the most inner) Andreev bound states $E_A^{+-}$ and $E_A^{-+}$ participate in the subgap charge transport. The frequency dependent Fano factor is then expressed by
\begin{equation}
\label{eq:flpm}
F_L(\omega) = 1 - \frac{4 \left( \alpha_-^4 + \alpha_+^4 \right) \Gamma_L^2}{9 \Gamma_L^2 + \omega^2} + \frac{10 \alpha_-^2 \alpha_+^2 \Gamma_L^2}{9 \Gamma_L^2 + \omega ^2} \; .
\end{equation}
The first (frequency-dependent) term comes from the auto-correlation tunneling processes through the Andreev bound states, for which the correlation functions are given by
\begin{equation}
\label{eq:sllpmpms}
S_{LL}^{+-,+-}(\omega) = \frac{4}{3} e^2 \alpha_-^2 \Gamma_L \left( 1 - \frac{4 \alpha_-^2 \Gamma_L^2}{9 \Gamma_L^2 + \omega^2} \right) \; ,
\end{equation}
\begin{equation}
\label{eq:sllmpmps}
S_{LL}^{-+,-+}(\omega) = \frac{4}{3} e^2 \alpha_+^2 \Gamma_L \left( 1 - \frac{4 \alpha_+^2 \Gamma_L^2}{9 \Gamma_L^2 + \omega^2} \right) \; .
\end{equation}
The second frequency-dependent term [appearing in Eq.~(\ref{eq:flpm})] comes from the electron-hole correlations
\begin{equation}
\label{eq:sllmppms}
S_{LL}^{+-,-+}(\omega) = \frac{20}{3} e^2 \frac{\alpha_-^2 \alpha_+^2 \Gamma_L^3}{9 \Gamma_L^2 + \omega^2} \; .
\end{equation}
Fig.~\ref{fig3}(b) displays the zero-frequency correlation function $S_{LL}$ and presents its components $S_{LL}^{+-,+-}$ and $S_{LL}^{-+,-+}$ obtained for $|\delta| < 1$ in the electron and hole transport through the Andreev bound states $E_A^{+-}$ and $E_A^{-+}$. We also present $S_{LL}^{+-,-+}$ describing inter-level correlations between the electron and hole tunneling events. In the conducting regime the main contribution to $S_{LL}$ comes from $S_{LL}^{+-,+-}$ and $S_{LL}^{-+,-+}$ which reduce the {zero-frequency} Fano factor $F \rightarrow 5/9$ in the extreme particle-hole asymmetry case (\textit{i.e.} a large $|\delta|$ limit). Notice, that such reduction of the zero-frequency Fano factor is smaller than in the N-QD-N system \cite{blanter2000, bulka2000}, where $F = 1/2$. This also follows from Eq.~(\ref{nqdn-1}) below for $\omega=0$ and $\Gamma_R=\Gamma_L$. Upon approaching the CB region we observe an enhancement of $S_{LL}$ due to activation of the inter-level correlations $S_{LL}^{+-,-+}$ (orange curve) corresponding to the DAR processes on L-QD interface. In the Coulomb blockade regime the current $J_L$ is exponentially suppressed and the corresponding frequency dependent Fano factor
\begin{equation}
\label{fano3tCB}
F_L(\omega) = 1 + 2 \alpha_-^2 - \frac{32 \alpha_-^4 \Gamma_L^2}{16 \Gamma_L^2 + \omega^2} + \frac{32 \alpha_-^2 \alpha_+^2 \Gamma_L^2}{16 \Gamma_L^2 + \omega^2} \; .
\end{equation}
The Schottky term is here enhanced by backscattering processes. The first and the second frequency-dependent terms refer to the auto-correlations and inter-level correlations, respectively. Right in a middle of the Coulomb blockade the frequency-dependent terms cancel each other, and $F_L(\omega = 0) = 2$ (due to backscattering).

Let us notice, that in the two-terminal N-QD-S case, the frequency dependent Fano factor for the conducting regime is given by
\begin{align}
\label{eqn.18}
F_{2t}(\omega) = 1 & - \frac{4 \alpha_-^2 \alpha_+^2 \Gamma_L^2}{(\alpha_-^2 + 2 \alpha_+^2)^2 \Gamma_L^2 + \omega^2} \nonumber \\
& + \frac{(\alpha_-^4 + 4 \alpha_+^4) \Gamma_L^2}{(\alpha_-^2 + 2 \alpha_+^2)^2 \Gamma_L^2 + \omega^2} \; .
\end{align}
The second frequency-dependent term [appearing in Eq.~(\ref{eqn.18})] corresponds to the inter-level correlations which is positive and has dominant contribution. Thereby $F_{2t}(\omega = 0) > 1$ and the noise has super-Poissonian character. It is always larger than in the three-terminal case -- compare the solid and dashed curve in Fig.~\ref{fig3}(a). In the Coulomb blockade regime the current and the noise are exponentially suppressed but the frequency dependent Fano factor is finite
\begin{equation}
F_{2t}(\omega) = 1 + \frac{\alpha_-^2}{\alpha_+^2} - \frac{4 \alpha_-^2 \Gamma_L^2}{\alpha_+^2 ( 4 \Gamma_L^2 + \omega^2 )} + \frac{4 \Gamma_L^2}{4 \Gamma_L^2 + \omega^2} \; .
\end{equation}
One can see, that inter-level correlations are responsible for the super-Poissonian noise -- compare with Eq.~(\ref{fano3tCB}). The positive correlations have been also found for multi-level QD coupled to the normal leads, where the super-Poissonian noise is driven by the inter-channel Coulomb blockade \cite{bulka1999}. In the regime of doubly occupied configuration, for the Coulomb blockade and for the empty state $F_L = 2$ due to uncorrelated jumps of the Cooper pairs to/from the proximized QD. In the conducting regime the dynamical fluctuations between DAR processes induce the negative correlations, one hence observes a reduction of $F_L$.

Fig.~\ref{fig3}(c) presents the zero-frequency cross-correlation function $S_{LR}$ between the different metallic leads. Let us remark, that in N-QD-N structures the charge conservation rule implies the relation $S_{LL} = -S_{LR}$ fulfilled at $\omega = 0$~\cite{blanter2000}. The function $S_{LR}$ is negative for entire range of the gate voltage $\delta$. Far away from the Coulomb blockade $S_{LR} \approx -S_{LL}$ so the components $S_{LR}^{+-,+-}$ and $S_{LR}^{-+,-+}$ become dominant because of correlations originating from the ET tunneling processes through metallic junction (see \textit{e.g.}~\cite{anantram1996}). For the unidirectional transport one can obtain the analytical expression for the frequency-dependent cross-correlation function. For the small bias voltage the transfer rates: $\Gamma_{L+}^{+-} = \alpha_-^2 \Gamma_L$, $\Gamma_{L+}^{-+} = \alpha_+^2 \Gamma_L$, $\Gamma_{R-}^{-+} = \alpha_+^2 \Gamma_R$, $\Gamma_{R-}^{+-} = \alpha_-^2 \Gamma_R$, $\Gamma_{L-}^{-+} = \Gamma_{L-}^{+-} = \Gamma_{R+}^{+-} = \Gamma_{R+}^{-+} = 0$.
\begin{equation}
S_{LR} = - \frac{4e^2}{3} \frac{(5 \alpha_+^4 + 5 \alpha_-^4 - 8 \alpha_-^2 \alpha_+^2) \Gamma_L^3}{9 \Gamma_L^2 + \omega^2}
\end{equation}
with its components
\begin{equation}
S_{LR}^{+-,+-}(\omega) = - \frac{20}{3} e^2 \frac{\alpha_-^4 \Gamma_L^3}{9 \Gamma_L^2 + \omega^2} \; ,
\end{equation}
\begin{equation}
S_{LR}^{-+,-+}(\omega) = - \frac{20}{3} e^2 \frac{\alpha_+^4 \Gamma_L^3}{9 \Gamma_L^2 + \omega^2} \; ,
\end{equation}
\begin{equation}
\label{eq:slrpmmp}
S_{LR}^{+-,-+}(\omega) = \frac{16}{3} e^2 \frac{\alpha_-^2 \alpha_+^2 \Gamma_L^3}{9 \Gamma_L^2 + \omega^2} \; .
\end{equation}
In the case considered here the superconducting electrode plays a crucial role, especially close to the Coulomb blockade region where the Andreev reflections become important, see Fig.~\ref{fig3}(c). The inter-level cross-correlation function $S_{LR}^{+-,-+}$ is positive and becomes then dominant. This is a fingerprint of the correlations between CAR processes~\cite{anantram1996}.

Since we have determined $S_{LL}=S_{RR}$ and $S_{LR}$ at $\omega = 0$, we can get the correlation function $S_{SS} = S_{LL} + 2 S_{LR} + S_{RR}$ describing correlations in the Cooper pair flow through S-QD interface. As could be expected, $S_{SS}$ is enhanced close the Coulomb blockade region. The direct and crossed Andreev reflection processes contribute equally to the correlation function $S_{SS}$ in the symmetric configuration. Notice, that the net current $J_S = 0$ but the corresponding noise $S_{SS}$ is large. Its value exceeds the ET contribution in $S_{LL}$ -- compare with the blue and green curves in Fig.~\ref{fig3}(b).

\subsection{\label{sec:large-bias} Current correlations for large bias}

\begin{figure*}[tbh] % Fig 4
\includegraphics[width=0.86\linewidth,clip]{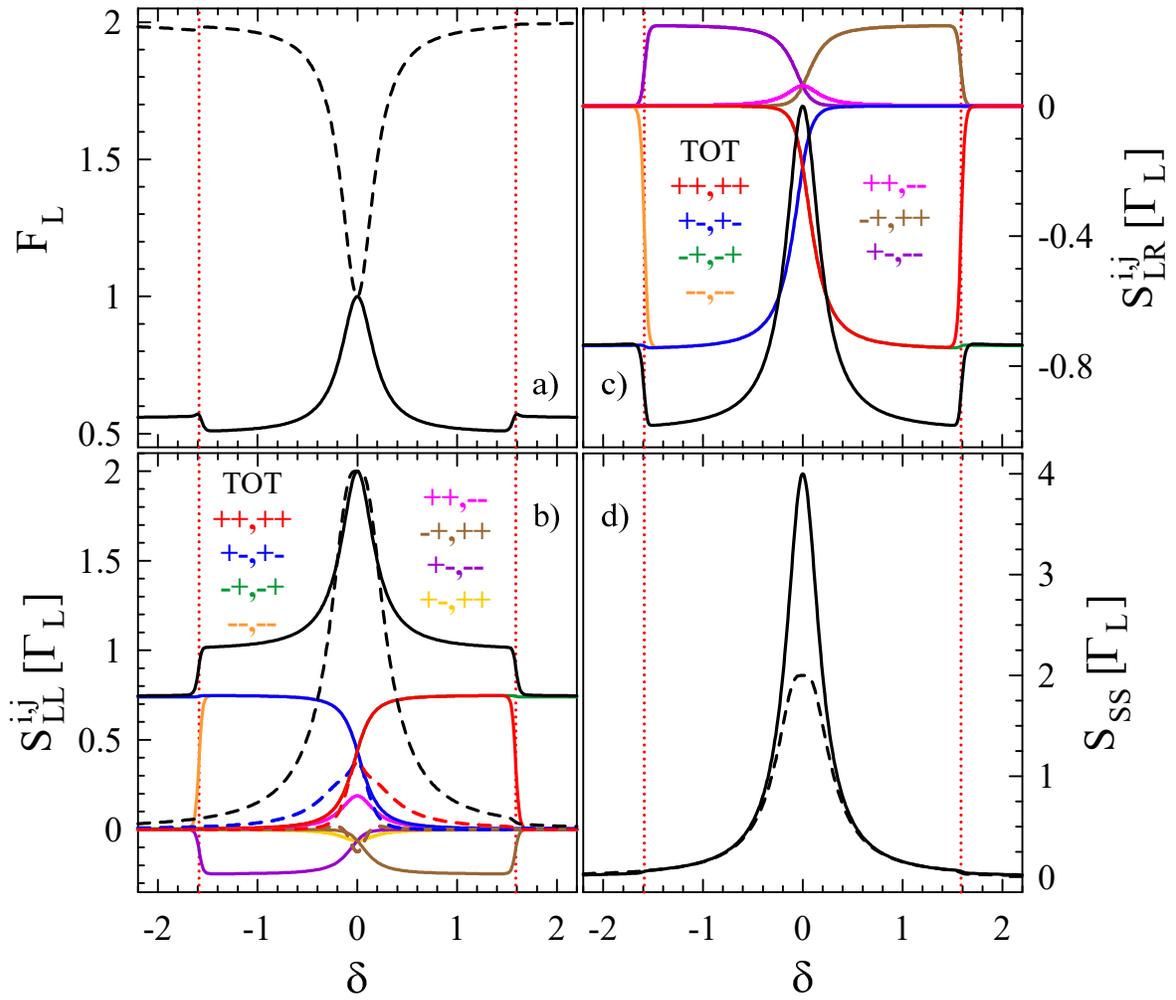}
\caption{The zero-frequency current correlation functions: (a) Fano factor $F_L$ of L-QD junction; (b) autocorrelation function $S_{LL}$ (black curve) and its contributions through various bound states (color curves); (c) cross-correlation function $S_{LR}$ (black curve) and its components. Note, that $S_{LR}^{+-,++}(\omega = 0) = S_{LR}^{++,--}(\omega = 0)$; and (d) autocorrelation function for $S$ electrode plotted as a function of gate voltage $\delta$ for moderate applied voltages $eV_L = 1.3 = - eV_R$ assuming the symmetric couplings $\Gamma_R = \Gamma_L$. For comparison the results for the two terminal case (N-QD-S) with $\Gamma_R = 0$ are presented by the dashed curves. Dotted red vertical lines denote positions of the Andreev bound state $E_A^{++}$. Other parameters are the same as those in Fig.~\ref{fig2}.} \label{fig4}
\end{figure*}

Now, let us consider the case of large bias voltage $eV_L=1.3=-eV_R$, when all the Andreev bound states participate in the subgap transport -- see Fig.~\ref{fig1}(b). The total current and its components are presented in Fig.~\ref{fig2}(b) in a moderate gate voltage range (for larger $|\delta|$ the QD is either in the double occupied or the empty state therefore the nanostructure becomes insulating). In the $|\delta| < E_A^{++}$ energy region one finds the current $J_L = e \Gamma_L$ and it is a sum of the appropriate electron and hole currents (see~Fig.~\ref{fig1}), obeying the conditions $J_L^{-+}= J_L^{++}=e \alpha_+^2 \Gamma_L/2$ and $J_L^{+-}= J_L^{--}=e \alpha_-^2 \Gamma_L/2$. Since we consider the symmetric case with $\Gamma_L=\Gamma_R$ the partial currents on R-QD junction are $J_R^{-+}= J_R^{++}=-e \alpha_+^2 \Gamma_L/2$ and $J_R^{+-}= J_R^{--}=-e \alpha_-^2 \Gamma_L/2$. This means absence of any charge accumulation on the Andreev bound states.

The current $J_L$ shows a step-like behavior, with large plateau for $|\delta| < E_A^{++}$. For large $|\delta|$ the total current is dominated by the ET processes. For $\delta = 0$ one finds that all electron and hole currents $J_L^i$ have the same amplitude, which indicates that DAR and ET processes equally participate in $J_L$. When $|\delta| > E_A^{++}$ the ET processes are seen only in $J_L^{+-}$ for $\delta < 0$ or $J_L^{-+}$ for $\delta > 0$. At $\delta = -E_A^{++}$ ($\delta = E_A^{++}$) the new transport channel opens (closes), therefore one observes an enhancement (a suppression) of the current $J_L^{--}$ ($J_L^{++}$) responsible for an enhancement (a suppression) of $J_L$. One can also find that $J_L^{--}(-\delta) = J_L^{++}(\delta)$ as well as $J_L^{+-}(-\delta) = J_L^{-+}(\delta)$, which is caused by the electron-hole symmetry. For comparison, in the two-terminal case, one observes the large peak (centered at $\delta = 0$) of $J_L$ due to DAR processes.

Let us analyze the results presented in Fig.~\ref{fig4} for the zero-frequency Fano factor $F_L$ and the current correlation functions $S_{LL}$, $S_{LR}$ and $S_{SS}$ at $\omega = 0$. One can notice, that the noise has sub-Poissonian character, with $F_L<1$, in the entire conducting range. The zero-frequency Fano factor is suppressed down to $1/2$ (for large $|\delta| > E_{++}$) due to the negative correlation in ET tunneling processes. In the centre of the plot, for $|\delta|\rightarrow 0$, we observe enhancement of the zero-frequency Fano factor caused by DAR processes. The inter-channel correlations become relevant, therefore $F_L \rightarrow 1$. We also performed the frequency-dependent calculations of the current correlations in the unidirectional transport regime. The frequency dependent Fano factor can be expressed as
\begin{equation}
F_L(\omega) = 1 - \frac{2 (\alpha_-^2 - \alpha_+^2)^2 \Gamma_L^2}{4 \Gamma_L^2 + \omega^2} \; .
\end{equation}
This result differs from that one, corresponding to the two terminal N-QD-S case
\begin{equation}\label{eq:fl2term}
F_{2t}(\omega) = 1 + \frac{(\alpha_-^2 - \alpha_+^2)^2 \Gamma_L^2}{\Gamma_L^2 + \omega^2} \; ,
\end{equation}
where the noise is super-Poisonian [compare the solid and dashed curves in Fig.~\ref{fig4}(a)]. The same result has been obtained using the diagrammatic real-time approach by Droste \textit{et al.}~\cite{droste2015} for the low-frequency dependent noise. It is also consistent with earlier studies by Braggio \textit{et al.}~\cite{braggio2011} who used the full counting statistics technique for the zero-frequency shot noise. The derivation of the current correlations and relaxation processes for the two-terminal case is outlined in the Supplementary Material~\cite{sm}.

Fig.~\ref{fig4}(b) shows various components of the correlation function $S_{LL}$. The frequency-dependent analysis of tunneling events shows important role of the intra- and inter-channel dynamics. Auto-correlation functions for electrons and holes can be written as
\begin{align}\label{eq:sllpppp}
S_{LL}^{++,++}(\omega) & = S_{LL}^{-+,-+}(\omega) \nonumber \\
& = e^2 \alpha_+^2 \Gamma_L \left( 1 - \frac{4 \alpha_+^2 \Gamma_L^2}{16 \Gamma_L^2 + \omega^2} \right) \; ,
\end{align}
\begin{align}
S_{LL}^{--,--}(\omega) & = S_{LL}^{+-,+-}(\omega) \nonumber \\
& = e^2 \alpha_-^2 \Gamma_L \left( 1 - \frac{4 \alpha_-^2 \Gamma_L^2}{16 \Gamma_L^2 + \omega^2} \right) \; .
\end{align}
The inter-level correlations between electron or hole tunneling processes are expressed by
\begin{align}\label{eq:sllmppp}
S_{LL}^{+-,--}(\omega) &= -\frac{2 e^2 \alpha_-^4 \Gamma_L^3}{4 \Gamma_L^2 + \omega^2} + \frac{4 e^2 \alpha_-^4 \Gamma_L^3}{16 \Gamma_L^2 + \omega^2} \; ,
\end{align}
\begin{align}
S_{LL}^{-+,++}(\omega) &= -\frac{2 e^2 \alpha_+^4 \Gamma_L^3}{4 \Gamma_L^2 + \omega^2} + \frac{4 e^2 \alpha_+^4 \Gamma_L^3}{16 \Gamma_L^2 + \omega^2} \;.
\end{align}
Notice that in these expressions the first frequency term is negative and it dominates at small frequencies, leading to sub-Poissonian noise. The electron-hole correlation functions between different channels are given by
\begin{align}
\label{eq:sllppmm}
S_{LL}^{++,--}(\omega) & = S_{LL}^{-+,+-}(\omega) \nonumber \\
& = \frac{2 e^2 \alpha_-^2 \alpha_+^2 \Gamma_L^3}{4 \Gamma_L^2 + \omega^2} + \frac{4 e^2 \alpha_-^2 \alpha_+^2 \Gamma_L^3}{16 \Gamma_L^2 + \omega^2} \; ,
\end{align}
\begin{equation}
\label{eq:sllpmpp}
S_{LL}^{+-,++}(\omega) = S_{LL}^{-+,--}(\omega) = -\frac{4 e^2 \alpha_-^2 \alpha_+^2 \Gamma_L^3}{16 \Gamma_L^2 + \omega^2} \; .
\end{equation}
The sum of all these components, Eqs.~(\ref{eq:sllpppp})-(\ref{eq:sllpmpp}), leads to
\begin{equation}\label{eq:sll}
S_{LL}(\omega) = 2 e^2 \Gamma_L \left[ 1 - \frac{2 (\alpha_-^2 - \alpha_+^2)^2 \Gamma_L^2}{4 \Gamma_L^2 + \omega^2} \right] \; ,
\end{equation}
where only the low-frequency fluctuations play a role whereas the contributions (negative and positive) of the high-frequency fluctuations cancel themselves.

Denominators of the frequency-dependent terms appearing in the correlation functions [Eqs.~(\ref{eq:sllpppp})-(\ref{eq:sll})] describe a relaxation driven by the generation-recombination processes \cite{vliet1965}. For any local quantity described by an operator $\hat{X}$ its dynamical fluctuation can be expressed by
\begin{align}
S_{XX}(\omega) = 4 \sum_{m, n} X_m G_{mn}(\omega) X_n p_n^0 \; ,
\label{eq:chp}
\end{align}
where $X_m$ is an eigenvalue of $\hat{X}$. Eqs.~(\ref{eq:scabij}) and (\ref{eq:chp}) are consistent with the quantum regression theorem, which predicts that the equation of motion for the statistically averaged operator implies the same equation for the time-correlation function of this operator~\cite{swain1981,breuer2002}.
Following~\cite{michalek2002} one can define operators for a local charge $\hat{N} = \hat{n}_+ + \hat{n}_-$ and polarization $\hat{P} = \hat{n}_+ - \hat{n}_-$, where $\hat{n}_+$ and $\hat{n}_-$ are the number operators for the state $|+\rangle$ and $|-\rangle$, respectively. In the case of large bias (considered here) the corresponding charge (polarization) noises are given by
\begin{align}
S_{NN} & = \frac{4 \Gamma_L}{16 \Gamma_L^2 + \omega^2} \; , \\
S_{PP} &= \frac{4 \Gamma_L}{4 \Gamma_L^2 + \omega^2} \; .
\end{align}
Thus, one can identify the terms with the relaxation rate $1/\tau_{rel}^{ch} = 4 \Gamma_L$ in the current noise as corresponding to the high-frequency charge fluctuations on the Andreev bound states, while the terms with $1/\tau_{rel}^{pol} = 2 \Gamma_L$ describe the low-frequency polarization fluctuations~\cite{bulka1999, michalek2002, bulka2008}.

Since we focus on the symmetric situation, the polarization fluctuations occur only in the total current noise $S_{LL}(\omega)$, Eq. (\ref{eq:sll}). When the symmetry was broken, for $\Gamma_L\neq \Gamma_R$ or for an asymmetric bias voltage, the displacement currents \cite{marcos2010, lavagna2018} as well as the accumulated charge \cite{bulka2000} could substantially affect the current noise, amplifying the frequency dependent Fano factor. In the asymmetric case $\Gamma_R \neq \Gamma_L$ the formulae are more complicated but for completeness and future reference we present them in the Supplementary Material~\cite{sm}.

Fig.~\ref{fig4}(c) presents the zero-frequency cross-correlation function $S_{LR}$ and its components between the metallic leads. This quantity is negative, indicating that ET tunneling processes and their correlations are dominant. At the electron-hole symmetry point, $\delta = 0$, where the Andreev scatterings become important, $S_{LR}$ tends to zero. This means that inter-electrode tunneling events are uncorrelated, \textit{i.e.} Poissonian. The frequency analysis of the current correlation gives
\begin{equation}
S_{LR}(\omega) = -\frac{4 e^2 (\alpha_-^2 - \alpha_+^2)^2 \Gamma_L^3}{4 \Gamma_L^2 + \omega^2}
\end{equation}
with the following components
\begin{align}
S_{LR}^{++,++}(\omega) & = S_{LR}^{-+,-+}(\omega) \nonumber \\
& = -\frac{2 e^2 \alpha_+^4 \Gamma_L^3}{4 \Gamma_L^2 + \omega^2} - \frac{4 e^2 \alpha_+^4 \Gamma_L^3}{16 \Gamma_L^2 + \omega^2} \; ,
\end{align}
\begin{align}
S_{LR}^{--,--}(\omega) & = S_{LR}^{+-,+-}(\omega) \nonumber \\
& = -\frac{2 e^2 \alpha_-^4 \Gamma_L^3}{4 \Gamma_L^2 + \omega^2} - \frac{4 e^2 \alpha_-^4 \Gamma_L^3}{16 \Gamma_L^2 + \omega^2} \; ,
\end{align}
\begin{equation}
S_{LR}^{-+,++}(\omega) = \frac{4 e^2 \alpha_+^4 \Gamma_L^3}{16 \Gamma_L^2 + \omega^2} \; ,
\end{equation}
\begin{equation}
S_{LR}^{+-,--}(\omega) = \frac{4 e^2 \alpha_-^4 \Gamma_L^3}{16 \Gamma_L^2 + \omega^2} \; ,
\end{equation}
\begin{equation}
S_{LR}^{++,--}(\omega) = S_{LR}^{-+,+-}(\omega) = \frac{4 e^2 \alpha_-^2 \alpha_+^2 \Gamma_L^3}{16 \Gamma_L^2 + \omega^2} \; ,
\end{equation}
\begin{align}
S_{LR}^{+-,++}(\omega) & = S_{LR}^{-+,--}(\omega) \nonumber \\
& = \frac{2 e^2 \alpha_-^2 \alpha_+^2 \Gamma_L^3}{4 \Gamma_L^2 + \omega^2} - \frac{4 e^2 \alpha_-^2 \alpha_+^2 \Gamma_L^3}{16 \Gamma_L^2 + \omega^2} \; .
\label{eq:slrpmpp}
\end{align}
Notice that $S_{LR}^{+-,++} (\omega) = S_{LR}^{-+,--} (\omega)$ [presented in Eq.~(\ref{eq:slrpmpp})] describes the electron-hole correlation between the different Andreev bound states and it is responsible for suppressing $S_{LR}(\omega = 0) \rightarrow 0$ in the limit $ \delta = 0$.

In analogy to the previous discussion, we observe that autocorrelation of the $J_{SS}$ currents in the S-QD junction of the three terminal system is strongly enhanced at $\delta = 0$, as displayed in Fig.~\ref{fig4}(d). This is caused by the frequency-dependent part, which is negative and vanishes at $\delta=0$, see Eq.~(\ref{eq:sll}). Its value is equal to the Schottky noise and is twice as large as in two terminal N-QD-S case, see dashed curve in Fig.~\ref{fig4}(d). This is the result of the aforementioned lack of correlation between electron tunneling through both normal tunnel junctions, therefore the noise simplifies to $S_{SS}(0) = 2 S_{LL}(0) = 4 e^2 \Gamma_L = 4 e J_L$.

\subsection{\label{sec:freq} Frequency dependence of the Fano factors: experimental consequences}

Till now we have been discussing the frequency dependence of the currents and their
fluctuations in the three terminal system with one of terminals being a superconductor.
To our knowledge there are no experimental data fitting our assumptions for such geometry.
Instead, the frequency dependent current statistics have been measured  experimentally
for a system consisting of two terminal quantum dot \cite{ubbelohde2012} with normal leads.
Our formalism is flexible enough to describe the experimentally studied system. Thus in the
following we shall compare the calculated frequency dependence of the Fano factors with
experimental data \cite{ubbelohde2012} obtained in the large bias limit.

The experimentally investigated single electron transistor \cite{ubbelohde2012} consisted of the interacting quantum
dot tunnel coupled to two normal electrodes. The authors~\cite{ubbelohde2012} have studied the frequency dependence of the second and third cumulants of the currents. In the Coulomb blockade regime the individual charge transport events were measured \textit{via} the quantum point contact placed nearby the quantum dot. The obtained current-current correlation function clearly shows the characteristic frequency dependence with a single correlation time.

\begin{figure}[tbh] % Fig 5
\includegraphics[width=0.86\linewidth,clip]{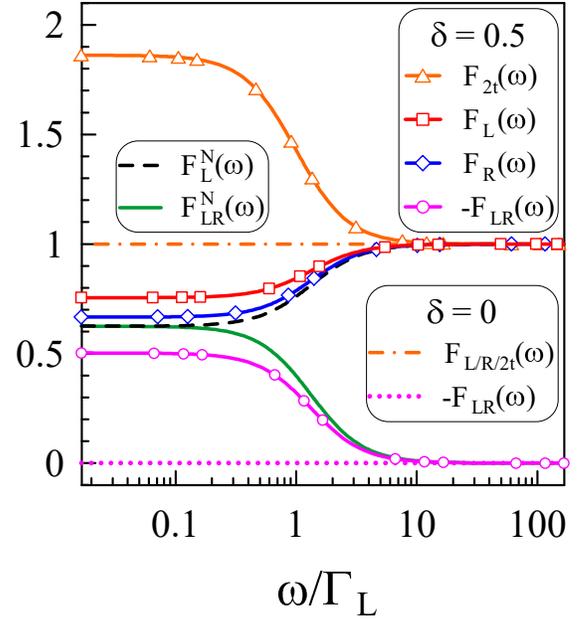}
\caption{{The frequency dependent Fano factors $F_L(\omega)$ [$F_R(\omega)$] of L-QD [QD-R] junction and $F_{LR}(\omega) = S_{LR}(\omega) / 2e \sqrt{| J_L J_R |}$ in three-terminal setup with the proximized quantum dot for the large bias voltage (see the Supplementary Material~\cite{sm}). The Fano factors $F_{2t}(\omega)$ [see Eq. (\ref{eq:fl2term})] of N-QD-S system are compared to $F_L^N(\omega)$ and $F_{LR}^N(\omega)$ of N-QD-N device, which has been investigated in Ref.~\cite{ubbelohde2012}. For the proximized dot in three-terminal setup our calculations have been performed using  $\Gamma_L = 3 \Gamma_R = 0.006$, $\Gamma_S = 0.2$, $k_BT = 0.01$ and $\delta = \{ 0, 0.5 \}$, while for the  dot hybridized with two normal electrodes we assumed $\epsilon_1 = 0$ ($U = 1$).}}
\label{fig5}
\end{figure}

Repeating the calculations for the quantum dot asymmetrically coupled to two normal electrodes we find the frequency dependent Fano factor
\begin{equation}
\label{nqdn-1}
F_R^N(\omega) = F_L^N(\omega) = 1 - \frac{2 \Gamma_L \Gamma_R}{( \Gamma_L + \Gamma_R)^2 + \omega^2} \; ,
\end{equation}
for the large bias processes, when the electrons tunnel in one direction only. The corresponding cross-correlation factor reads (up to the sign, as in the main text we defined the cross-correlations with a minus sign)
\begin{equation}
\label{nqdn-2}
F_{LR}^N(\omega) \equiv \frac{S_{LR}^N(\omega)}{2e J} = \frac{\Gamma_L^2 + \Gamma_R^2}{( \Gamma_L + \Gamma_R )^2 + \omega^2} \; .
\end{equation}

In the experiment the couplings where strongly asymmetric with $\Gamma_L \approx 3 \Gamma_R$ and the data, shown in Fig.~1d of the experimental paper~\cite{ubbelohde2012}, clearly feature single relaxation rate $1/\tau_c=\Gamma_L+\Gamma_R$. The experimental data essentially overlap the theoretical curves marked as $F_L^N(\omega)$ and $F_{LR}^N(\omega)$ in Fig.~\ref{fig5}. This figure not only illustrates the frequency dependence of the factors (\ref{nqdn-1}) and (\ref{nqdn-2}), but also shows the corresponding frequency dependent Fano factors of our two and three terminal proximized quantum dot. For the latter system we have defined these frequency dependent factors as $F_L(\omega) = S_{LL}(\omega) / 2eJ_L$, $F_R(\omega) = S_{RR}(\omega) / 2eJ_R$ and $F_{LR}(\omega) = S_{LR}(\omega) / 2e \sqrt{|J_L J_R|}$ (see the Supplementary Material~\cite{sm}) for two values of $\delta = \{0, 0.5 \}$ as indicated. In the three terminal hybrid system the largest effect of the superconducting electrode (superconducting correlation) is seen for the electron-hole symmetry point ($\delta = 0$). As $\delta$ increases, the effect of the superconducting electrode clearly diminishes while for the N-QD-S system the effect increases with $\delta$. The largest discrepancies between the two- or three-terminal hybrid and the normal (two terminal) system occur for the low frequencies. If the hybrid system is close to the charge degeneracy point $\delta\approx 0$ the contribution to $S_{LR}(\omega)$ identically vanishes as also does the frequency dependent part of $F_L(\omega)$ and $F_{2t}(\omega)$, so it attains the form of Schottky noise with $F_L = F_{2t} = 1$ independently of frequency. In Fig.~\ref{fig5} we also show the comparison of the frequency dependent Fano factors for two normal terminals contacting the quantum dot when the Coulomb blockade is expected to play an important role with those of the three terminal proximized dot as well as N-QD-S system. One can see that the Andreev scattering makes all frequency dependent Fano factors to attain different values for $\omega \rightarrow 0$. This is related to charge non-conserving processes in individual normal electrodes. For $\delta = 2$ (not shown) the frequency dependent Fano factors for the three terminal dot with the superconducting electrode nearly coincide with those obtained for the system without proximity effect. This is traced back to the weak mixing of the empty and doubly occupied states for values of $\delta \gg \Gamma_S$.

On the other hand for N-QD-S system the frequency dependent Fano factor $F_{2t}(\omega)$ for large $\delta$ approaches $2$ which clearly indicates that the Cooper pairs tunneling is a Poissonian process. One can also easily check that with increasing asymmetry between $\Gamma_L$ and $\Gamma_R$ (\textit{i.e.} with decrease of $\Gamma_R$) one can observe that $F_L(\omega)$ grows from the sub-Poissonian to the super-Poissonian values and in the limit $\Gamma_R \rightarrow 0$ one finds $F_L(\omega) = F_{2t}(\omega)$.

\section{\label{sec:summary} Summary of main findings}

We have studied the charge current fluctuations for three-terminal hybrid system, comprising the QD sandwiched between two metallic leads and strongly coupled to superconductor. Using the generation-recombination approach~\cite{korotkov1994, vliet1965} we have considered electron and hole tunneling through available subgap channels (in-gap bound states) of such proximized QD. In particular, we have addressed mutual relationship between the single electron transport (ET) and the Andreev (electron to hole) scattering caused by the local (DAR) and non-local (CAR) mechanisms. We have determined the currents and identified their components originating from transport through the specific bound states. We have analyzed the current-current correlation functions and performed their spectral decompositions, getting insight into the local and non-local fluctuations responsible for suppression or enhancement of the current noise. We have done numerical computations, focusing on the small and large bias voltages that activate either two or four in-gap Andreev levels in the transport window, respectively.

Assuming the symmetric couplings $\Gamma_R = \Gamma_L$ and symmetric bias $V_L = V / 2 = - V_R$ we have found, that single electron transport (ET) processes play important role in the local current-current correlation $S_{LL}$, leading to reduction of the shot noise to the sub-Poissonian regime with the zero-frequency Fano factor $F_L < 1$.
Deviation from the Poissonian noise is caused by the dynamical (frequency dependent) part of the current correlations. Its intra-channel components (tunneling processes through the same Andreev bound state) bring always the negative contribution, manifesting the anti-bunching behavior. On the other hand, the inter-channel components can bring positive contribution. For small bias there appear positive correlations between the electrons and holes, indicating that the currents to the same normal lead originate from the direct Andreev reflection (DAR) processes. In such situation the intra- and inter-channel components describe the charge fluctuations with the same relaxation rate. On the other hand, for large bias, we observe two different charge fluctuation processes: one with the large relaxation rate $1/\tau_{rel}^{ch} = 4 \Gamma_L$ and another with the small relaxation rate $1/\tau_{rel}^{pol} = 2 \Gamma_L$, respectively. Since in the symmetric case the charge is not accumulated on the Andreev bound states, the current correlation functions reveal a clear contribution from the charge and polarization fluctuations. The intra-channel correlations show up only the negative component, corresponding to the high frequency charge relaxation, whereas the inter-channel correlations have both the frequency terms. DAR processes are well manifested in the inter-channel functions $S_{LL}^{++,--}(\omega)$ and $S_{LL}^{-+,+-}(\omega)$ [see Eq.~(\ref{eq:sllppmm})] with both positive frequency-dependent contributions. We noticed a somewhat intriguing behavior in charge fluctuations, where the intra and inter-channel correlation components compensate each other. For this reason the correlation function $S_{LL}(\omega)$ for the total current has the frequency-dependent contribution, originating solely from the polarization fluctuations.

To emphasize the role of ET processes we have contrasted our results with those obtained for the two-terminal N-QD-S case, where the crossed Andreev reflection (CAR) processes are absent. Under such circumstances the bunching effect is well seen in the electron-hole correlation functions, implying the super-Poissonian noise with the zero-frequency Fano factor $F_{2t} > 1$. Our studies provide in-depth information about the dynamics of various internal charging processes. In particular, we can see relationship between the low-frequency polarisation fluctuations and the high-frequency charge fluctuations in all components of the current correlation functions. This substantially extends the previous studies by Droste~\cite{droste2015} and Braggio~\cite{braggio2011}.

For the three-terminal heterostructure we have also inspected the nonlocal correlations between currents flowing through the left and the right tunnel junctions to metallic electrodes. The corresponding cross-correlation function $S_{LR}$ exhibits a nontrivial interplay between the CAR and ET processes. For various gate voltages (arbitrary detuning $\delta$ from the half-filled QD) we have found the negative values of $S_{LR}$, which indicates antibunching of the tunneling events typical for multi-terminal normal systems (see \textit{e.g.}~\cite{bulka2008}). Although the ET processes play the dominant role, the nonlocal CAR processes are important, especially near the electron-hole symmetry point, $|\delta|\rightarrow 0$, where $S_{LR}\rightarrow 0$. Enhancement of the shot noise originates from activation of the inter-channel fluctuations between electron and holes. In the case of small bias this is due to the correlation function $S_{LR}^{+-,-+}(\omega)$ [Eq.~(\ref{eq:slrpmmp})] contributed by electrons and holes from different leads which could be related to the CAR processes. For the large bias the correlation functions $S_{LR}^{+-,++}(\omega)$ and $S_{LR}^{-+,--}(\omega)$ [Eq.~(\ref{eq:slrpmpp})], where the low frequency polarization fluctuations are positive, become more relevant.

\section{\label{sec:concl-outl} Conclusions and outlook}

Our theoretical investigation of the quantum dot embedded into three-terminal heterostructure is quite universal and it yields proper expressions for the charge fluctuations and the frequency dependent Fano factors describing the statistical properties of the two terminal N-QD-S and N-QD-N systems. Quantitative agreement of the present results with experimental data obtained recently for N-QD-N setup~\cite{ubbelohde2012} and satisfactory agreement with the previous theoretical work on N-QD-S system~\cite{droste2015} gives a confidence that our method would be able to correctly describe the statistical correlations of the currents in hybrid systems with one superconducting and one or two normal electrodes. We thus hope that present analysis of the statistical properties of currents could stimulate further (experimental and theoretical) activities for this interesting geometry. In particular, our predictions can be verified experimentally in the system similar to that used already by Ubbelohde \textit{et al.}~[\onlinecite{ubbelohde2012}].

There is plenty of room for extending our studies in other directions as well. For instance, the auto- and cross-correlation have been discussed for the proximized QD coupled to magnetic electrodes in the Cooper pair splitter (CPS) configuration~\cite{weiss2017,trocha2018}. Such considerations, however, have been limited mostly to the zero-frequency case.
It has to be mentioned that we are not considering the capacitive effects, which can be of importance in particular experiments and affect the frequency dependence of correlation functions.
Our method, suitable for frequency dependent current-current correlations, seems to give a broader insight into the dynamical processes of the Cooper pair formation (or splitting) in such devices~\cite{michalek2020}. Influence of the short-time fluctuations on efficiency of the Cooper pair entanglement might be crucial for future application of the CPS \textit{e.g.} to quantum computation and/or communication. Furthermore, it has been recently observed, that the high-frequency cut-off completely washes out the emission noise related to the Kondo resonance~\cite{delagrange2018}. It would be hence challenging to check, whether such behavior can be overcome in the correlated quantum dots proximitized to superconductors, where the subgap Kondo peak is expected to be substantially broadened upon approaching the doublet-singlet quantum phase transition \cite{zitko2015, domanski2016}.

Another interesting perspective would be possible in nanostructures, where the superconducting lead is in a topologically nontrivial phase hosting the Majorana boundary modes. Some authors~\cite{wang2013} have pointed out, that non-local nature of these zero-energy modes could by manifested by the strong cross-correlations detectable in the shot noise measurements. Leakage of these boundary modes onto side-attached quantum dots has been also proposed as a suitable tool for unambiguous recognition of the Majorana zero-energy quasiparticles from their trivial (finite-energy) counterparts~\cite{lutchyn2015, flensberg2018}.

Furthermore, three-terminal junctions with the quantum dot sandwiched between the topologically nontrivial superconducting nanowires could induce the giant shot noise~\cite{jonckheere2019}. Other setup, with the topological superconducting island coupled to three normal-conducting leads, has been recently proposed~\cite{flensberg2019} for efficient protocol to test nonlocality of the Majorana bound states through the current shot-noise correlations, where the zero-frequency Fano factor could detect the exotic (nonabelian) character of the Majorana quasiparticles~\cite{lu2020}. These  examples show a rich variety of heterostructures, where our study can be potentially extended.

\appendix

\section{\label{appA} Master equation approach}

In the weak coupling limit $\Gamma_{L(R)} \ll k_BT, \Gamma_S$ transport properties of our nanostructure are dominated by a sequential tunneling processes through the proximized QD. The currents can be determined, by solving the master equation
\begin{equation}
\label{eq:master}
\mathbf{\dot{p}}(t) = \mathbf{M \; p}(t)
\end{equation}
with the probabilities $\mathbf{p}(t) = ( p_1(t), p_-(t), p_+(t) )^T$ referring to the single electron occupancy of QD $p_1(t) = p_\uparrow(t) + p_\downarrow(t)$ and probabilities of the BCS-type configurations $| - \rangle$ and $| + \rangle$ denoted by $p_-(t)$ and $p_+(t)$, respectively. Probability conservation implies the constraint $p_1(t) + p_-(t) + p_+(t) \equiv 1$.

The evolution matrix $\mathbf{M}$ has in our problem the following structure
\begin{equation}
\label{eq:matrix}
\mathbf{M} =
\left(
\begin{array}{ccc}
-M_{21} - M_{31} & M_{12} & M_{13} \\
M_{21} & -M_{12} & 0 \\
M_{31} & 0 & -M_{13} \\
\end{array}
\right) \; ,
\end{equation}
where $M_{12} = 2 \sum_{\alpha = L, R} ( \Gamma_{\alpha +}^{-+} + \Gamma_{\alpha -}^{+-} )$, $M_{13} = 2 \sum_{\alpha = L, R} ( \Gamma_{\alpha +}^{--} + \Gamma_{\alpha -}^{++} )$, $M_{21} = \sum_{\alpha = L, R} ( \Gamma_{\alpha +}^{+-} + \Gamma_{\alpha -}^{-+} )$ and $M_{31} = \sum_{\alpha = L, R} ( \Gamma_{\alpha +}^{++} + \Gamma_{\alpha -}^{--} )$. We have introduced here the effective transition rates $\Gamma_{\alpha \pm}^i$ describing tunneling processes to/from the QD ($+/-$) through $\alpha$-th junction and engaging the Andreev bound states $i \in \{ ABS \} \equiv \{ ++, +-, -+, -- \}$, see Fig.~\ref{fig1}.
The tunneling rates describe transfer of one electron or hole between the singly occupied states and the singlet subspace.
In particular, $\Gamma_{\alpha \pm}^{++} = \alpha_+^2 \Gamma_\alpha f(\pm E_A^{++} \mp \mu_\alpha)$, $\Gamma_{\alpha \pm}^{+-} = \alpha_-^2 \Gamma_\alpha f(\pm E_A^{+-} \mp \mu_\alpha)$, $\Gamma_{\alpha \pm}^{-+} = \alpha_+^2 \Gamma_\alpha f(\pm E_A^{-+} \mp \mu_\alpha)$ and $\Gamma_{\alpha \pm}^{--} = \alpha_-^2 \Gamma_\alpha f(\pm E_A^{--} \mp \mu_\alpha)$, where $f(E) = \left[ 1 + \exp{\left( E/k_{B}T \right)} \right]^{-1}$ is the Fermi-Dirac distribution function. In the case of positive Andreev bound states, $E_A^i > \mu_S$, the tunneling rate $\Gamma_{\alpha \pm}^i$ describes an electron transfer, whereas for $E_A^i < \mu_S$ the charge current is contributed by holes. The chemical potential of the superconducting electrode is denoted as $\mu_S$ and its value is assumed to be $\mu_S=0$ throughout.

The charge current flowing from $\alpha$-th lead can be expressed in the stationary limit as $J_\alpha = \sum_{i \in \{ ABS \}} J_\alpha^i$, where the contributions from the Andreev bound states are given by
\begin{equation}
\label{eq:currpp}
J_\alpha^{++} \equiv I_{\alpha +}^{++} - I_{\alpha -}^{++} = e ( \Gamma_{\alpha +}^{++} p_1^0 - 2 \Gamma_{\alpha -}^{++} p_+^0 ) \; ,
\end{equation}
\begin{equation}
\label{eq:currpm}
J_\alpha^{+-} \equiv I_{\alpha +}^{+-} - I_{\alpha -}^{+-} = e ( \Gamma_{\alpha +}^{+-} p_1^0 - 2 \Gamma_{\alpha -}^{+-} p_-^0 ) \; ,
\end{equation}
\begin{equation}
\label{eq:currmp}
J_\alpha^{-+} \equiv I_{\alpha +}^{-+} - I_{\alpha -}^{-+} = -e ( \Gamma_{\alpha -}^{-+} p_1^0 - 2 \Gamma_{\alpha +}^{-+} p_-^0 ) \; ,
\end{equation}
\begin{equation}
\label{eq:currmm}
J_\alpha^{--} \equiv I_{\alpha +}^{--} - I_{\alpha -}^{--} = -e ( \Gamma_{\alpha -}^{--} p_1^0 - 2 \Gamma_{\alpha +}^{--} p_+^0 ) \; .
\end{equation}
Under stationary conditions the probability $\mathbf{p}^0 = ( p_1^0, p_-^0, p_+^0 )^T$ can be determined, solving the equation $\mathbf{M p}^0 = \mathbf{0}$. The currents $J_\alpha^{++}$ and $J_\alpha^{+-}$ are contributed by electrons (`e'), whereas $J_\alpha^{-+}$ and $J_\alpha^{--}$ by holes (`h') \textit{via} the corresponding Andreev bound states, see Fig.~\ref{fig1}. One can notice, that for $| \delta | = 1$ the Andreev bound states $E_A^{+-}$ and $E_A^{-+}$ cross each other, signifying the quantum phase transition~\cite{hewson2007}. For large $| \delta | > 1$ the currents $J_\alpha^{+-}$ and $J_\alpha^{-+}$ describe hence transport of holes and electrons, respectively. As far as the Copper pair current $J_S$ is concerned (flowing from the QD to superconducting lead) it can be obtained from the Kirchoff's law $J_S = J_L + J_R$.

\begin{acknowledgments}
This research was financed by National Science Centre (Poland) under the project numbers 2016/21/B/ST3/02160 (BRB, GM) and 2017/27/B/ST3/01911 (TD, KIW).
\end{acknowledgments}

\end{document}

\title{Statistical correlations of currents flowing \textit{via} proximized quantum dot\\-- Supplementary Material --}

\author{G. Micha{\l}ek} \email{grzechal@ifmpan.poznan.pl}
\author{B. R. Bu{\l}ka}
\affiliation{Institute of Molecular Physics, Polish Academy of Sciences, ul. M. Smoluchowskiego 17, 60-179 Pozna\'{n}, Poland}

\author{T. Doma\'{n}ski}
\author{K. I. Wysoki\'{n}ski}
\affiliation{Institute of Physics, M. Curie-Sk{\l}odowska University, pl. M. Curie-Sk{\l}odowskiej 1, 20-031 Lublin, Poland}

\date{\today}

\begin{abstract}
Here we present some details concerning current correlations of the two terminal N - QD - S system as well
as three terminal system with arbitrary couplings $\Gamma_L$ and $\Gamma_R$ of the quantum dot
to two normal electrodes. The formulae used  in our article \cite{mainpaper} were specialised to
symmetric couplings with $\Gamma_L = \Gamma_R$.
\end{abstract}

%\pacs{74.45.+c, 73.23.Hk, 72.70.+m, 73.63.Kv}
% 74.45.+c Proximity effects; Andreev reflection; SN and SNS junctions
% 73.23.Hk Coulomb blockade; single-electron tunneling
% 72.70.+m Noise processes and phenomena
% 73.63.Kv Quantum dots

\maketitle

\section{Introduction}
This supplementary material contains the detailed formulae for the frequency dependent currents correlations
derived withing the approach which makes use of master equation in the
formulation of the paper \cite{korotkov1994}. The emphasis is
on the two terminal hybrid system consisting of the quantum dot coupled to one normal and one
superconducting electrode (Section \ref{appB}) and the hybrid system with one superconducting and two
normal electrodes (Section \ref{appC}).

\section{\label{appB} Frequency dependent noise of two-terminal N-QD-S case}

In the small bias voltage regime the states $E_A^{+-}$ and $E_A^{-+}$ only participate in transport. Here, we consider the unidirectional case in which the transition rates: $\Gamma_{L+}^{+-} = \alpha_-^2 \Gamma_L$, $\Gamma_{L+}^{-+} = \alpha_+^2\Gamma_L $, $\Gamma_{L-}^{+-} =\Gamma_{L-}^{-+} = 0$ and the current $J_{2t} = 4 e \alpha_-^2 \alpha_+^2 \Gamma_L / (\alpha_-^2 + 2 \alpha_+^2)$. The frequency dependent Fano factor can be written as
\begin{align}
F_{2t}(\omega) = 1 & - \frac{4 \alpha_-^2 \alpha_+^2 \Gamma_L^2}{\omega^2 + (\alpha_-^2 + 2 \alpha_+^2)^2 \Gamma_L^2} \nonumber \\
& + \frac{ (\alpha_-^4 + 4 \alpha_+^4) \Gamma_L^2}{\omega^2 + (\alpha_-^2 + 2 \alpha_+^2)^2 \Gamma_L^2} \;.
\end{align}
The first frequency dependent term comes from the auto-correlation tunneling processes through the level $E_A^{+-}$ and $E_A^{-+}$, and it is negative thus reducing the noise. The second term is due to inter-level current correlations, it dominates and leads to the super-Poissonian noise with $F_{2t}(\omega = 0) \rightarrow 2$ for large $|\delta|$. The relaxation rate $1 / \tau_{rel} = ( \alpha_-^2 + 2 \alpha_+^2) \Gamma_L$ is related with a generation and recombination process~\cite{vliet1965} for the Andreev scattering.

In the Coulomb blockade regime $\Gamma_{L+}^{+-} = \alpha_-^2 \Gamma_L f(E_A^{+-})$ is exponentially suppressed as well as the current $J_{2t} \simeq 2 e \alpha_- \alpha_+ \Gamma_L f(E_A^{+-})$, whereas the frequency dependent Fano factor is given by
\begin{equation}
F_{2t}(\omega) = 1 + \frac{\alpha_-^2}{\alpha_+^2} - \frac{4 \alpha_-^2 \Gamma_L^2}{\alpha_+^2 ( \omega^2 + 4 \Gamma_L^2 )} + \frac{4 \Gamma_L^2}{\omega^2 + 4 \Gamma_L^2} \; .
\end{equation}

For a large bias voltage, when all Andreev bound states participate in transport and in the unidirectional limit the rates $\Gamma_{L +}^{++} = \alpha_+^2 \Gamma_L$, $\Gamma_{L +}^{--} = \alpha_-^2 \Gamma_L$, $\Gamma_{L +}^{+-} = \alpha_-^2 \Gamma_L$, $\Gamma_{L +}^{-+} = \alpha_+^2 \Gamma_L$, and the others are equal zero. The currents flowing through the Andreev bound states are
\begin{equation}
J_{2t}^{+-} = J_L^{-+} = 2 e \alpha_-^4 \alpha_+^2 \Gamma_L \; ,
\end{equation}
\begin{equation}
J_{2t}^{--} = J_L^{++} = 2 e \alpha_-^2 \alpha_+^4 \Gamma_L \; .
\end{equation}
The frequency dependent Fano factor
\begin{equation} \label{eq:F2t}
F_{2t}(\omega) = 1 + \frac{\Gamma_L^2 \Delta_\alpha^2}{\omega^2 + \Gamma_L^2}
\end{equation}
and the current auto-correlation functions are expressed as
\begin{align}
& S_{2t}^{+-,+-}(\omega) = S^{-+,-+}(\omega) = \nonumber \\
& 4 e^2 \alpha_-^4 \alpha_+^2 \Gamma_L \left( 1 - \frac{4 \alpha_-^2 \alpha_+^2 \Gamma_L^2 \Delta_\alpha }{\omega^2 + \Gamma_L^2} - \frac{8 \alpha_-^2 \alpha_+^4 \Gamma_L^2}{\omega^2 + 4 \Gamma_L^2} \right) \; ,
\end{align}
\begin{align}
& S_{2t}^{--,--}(\omega) = S^{++,++}(\omega) = \nonumber \\
& 4 e^2 \alpha_-^2 \alpha_+^4 \Gamma_L \left( 1 + \frac{4 \alpha_-^2 \alpha_+^2 \Gamma_L^2 \Delta_\alpha}{\omega^2 + \Gamma_L^2} - \frac{8 \alpha_-^4 \alpha_+^2 \Gamma_L^2}{\omega^2 + 4 \Gamma_L^2} \right) \; ,
\end{align}
while the cross-correlation functions are
\begin{align}
S_{2t}^{-+,+-}(\omega) & = \frac{4 e^2 \alpha_-^6 \alpha_+^2 (\alpha_-^4 - 2 \alpha_-^2 \alpha_+^2 + 5 \alpha_+^4) \Gamma_L^3}{\omega^2 + \Gamma_L^2} \nonumber \\
& + \frac{16 e^2 \alpha_-^4 \alpha_+^4 (\alpha_-^4 + \alpha_+^4) \Gamma_L^3}{\omega^2 + 4 \Gamma_L^2} \; ,
\end{align}
\begin{align}
S_{2t}^{++,--}(\omega) & = \frac{4 e^2 \alpha_-^2 \alpha_+^6 (5\alpha_-^4 - 2 \alpha_-^2 \alpha_+^2 + \alpha_+^4) \Gamma_L^3}{\omega^2 + \Gamma_L^2} \nonumber \\
& + \frac{16 e^2 \alpha_-^4 \alpha_+^4 (\alpha_-^4 + \alpha_+^4) \Gamma_L^3}{\omega^2 + 4 \Gamma_L^2} \; ,
\end{align}
\begin{align}
& S_{2t}^{+-,--}(\omega) = S^{-+,++}(\omega) = \nonumber \\
& 4 e^2 \alpha_-^4 \alpha_+^4 \Gamma_L^3 \left( \frac{\alpha_-^4 - 6 \alpha_-^2 \alpha_+^2 + \alpha_+^4}{\omega^2 + \Gamma_L^2} + \frac{8 \alpha_-^2 \alpha_+^2}{\omega^2 + 4 \Gamma_L^2} \right) \; ,
\end{align}
\begin{align}
& S_{2t}^{+-,++}(\omega) = S^{-+,--}(\omega) = \nonumber \\
& 8 e^2 \alpha_-^4 \alpha_+^4 \Gamma_L^3 \left[ \frac{\Delta_\alpha^2}{\omega^2 + \Gamma_L^2} - \frac{2 (\alpha_-^4 + \alpha_+^4)}{\omega^2 + 4 \Gamma_L^2} \right] \; ,
\end{align}
where $\Delta_\alpha = \alpha_-^2 - \alpha_+^2$. Thus the total noise is
\begin{equation}
S_{2t} = 8 e^2 \alpha_-^2 \alpha_+^2 \Gamma_L \left[ 1 + \frac{\Gamma_L^2 \Delta_\alpha^2}{\omega^2 + \Gamma_L^2} \right] \; .
\end{equation}
Notice that there is only the low frequency term, with the relaxation rate $1 / \tau_{rel}^{pol} = \Gamma_L$ describing the polarization fluctuations, while the high frequency terms with $1 / \tau_{rel}^{ch} = 2 \Gamma_L$ cancel each other.

\section{\label{appC} Frequency dependent noise for three-terminal case: role of asymmetry}

Here, we consider the general case of the unidirectional transport with the asymmetrical coupling $\Gamma_L \neq \Gamma_R$ and the symmetrical bias voltage, $\mu_L = e V / 2$, $\mu_R = - e V / 2$.

For the {\bf small bias voltage} the transfer rates: $\Gamma_{L+}^{+-} = \alpha_-^2 \Gamma_L$, $\Gamma_{L+}^{-+} = \alpha_+^2 \Gamma_L$, $\Gamma_{R-}^{-+} = \alpha_+^2 \Gamma_R$, $\Gamma_{R-}^{+-} = \alpha_-^2 \Gamma_R$, $\Gamma_{L-}^{-+} = \Gamma_{L-}^{+-} = \Gamma_{R+}^{+-} = \Gamma_{R+}^{-+} = 0$. The current flowing through the L junction can be expressed as $J_L = J_L^{-+} + J_L^{+-}$, where its components are
\begin{equation}
J_L^{-+} = \frac{2 e \alpha_+^2 \Gamma_L z_2}{2 z_1 + z_2} \; ,
\end{equation}
\begin{equation}
J_L^{+-} = \frac{2 e \alpha_-^2 \Gamma_L z_1}{2 z_1 + z_2} \; .
\end{equation}
Here, we denoted by $z_1 = \alpha_+^2 \Gamma_L + \alpha_-^2 \Gamma _R$ and $z_2 = \alpha_-^2 \Gamma_L + \alpha_+^2 \Gamma _R$.

The current correlation functions are expressed as
\begin{align}
S_{LL}^{-+,-+}(\omega) & = \frac{4 e^2 \alpha_+^2 \Gamma_L z_2}{2 z_1 + z_2} \nonumber \\
& - \frac{16 e^2 \alpha_+^4 \Gamma_L^2}{2 z_1 + z_2} \frac{z_2^2}{\omega^2 + (2 z_1 + z_2)^2} \; ,
\end{align}
\begin{align}
S_{LL}^{+-,+-}(\omega) & = \frac{4 e^2 \alpha_-^2 \Gamma_L z_1}{2 z_1 + z_2} \nonumber \\
& - \frac{16 e^2 \alpha_-^4 \Gamma_L^2}{2 z_1 + z_2} \frac{z_1^2}{\omega^2 + (2 z_1 + z_2)^2} \; ,
\end{align}
\begin{equation}
S_{LL}^{-+,+-}(\omega) = \frac{4 e^2 \alpha_-^2 \alpha_+^2 \Gamma_L^2}{2 z_1 + z_2} \frac{4 z_1^2 + z_2^2}{\omega^2 + (2 z_1 + z_2)^2} \; .
\end{equation}
The corresponding frequency dependent Fano factor
\begin{align}
F_L(\omega) = 1 + \frac{2 \Gamma_L}{\alpha_-^2 z_1 + \alpha_+^2 z_2} & \left[ - \frac{2 ( \alpha_-^4 z_1^2 + \alpha_+^4 z_2^2 )}{\omega^2 + (2 z_1 + z_2)^2} \right. \nonumber \\
& + \left. \frac{\alpha_-^2 \alpha_+^2 \left(4 z_1^2 + z_2^2\right)}{\omega^2 + (2 z_1 + z_2)^2} \right] \; ,
\end{align}
where the first/second frequency dependent term comes from intra/inter channel correlations.

The cross correlation function $S_{LR}(\omega)$ and its components are
\begin{align}
S_{LR}(\omega) = \frac{4 e^2 \Gamma _L \Gamma _R}{2 z_1 + z_2} & \left[ \frac{4 \alpha_-^2 \alpha_+^2 ( z_1^2 + z_2^2 )}{\omega^2 + ( 2 z_1 + z_2)^2} \right. \nonumber \\
& - \left. \frac{( \alpha_-^4 + \alpha_+^4 ) ( 4 z_1^2 + z_2^2 )}{\omega^2 + ( 2 z_1 + z_2)^2} \right] \; ,
\end{align}
\begin{equation}
S_{LR}^{-+,-+}(\omega) = - \frac{4 e^2 \alpha_+^4 \Gamma_L \Gamma_R}{2 z_1 + z_2} \frac{4 z_1^2 + z_2^2}{\omega^2 + (2 z_1 + z_2)^2} \; ,
\end{equation}
\begin{equation}
S_{LR}^{+-,+-}(\omega) = - \frac{4 e^2 \alpha_-^4 \Gamma_L \Gamma_R}{2 z_1 + z_2} \frac{4 z_1^2 + z_2^2}{\omega^2 + (2 z_1 + z_2)^2} \; ,
\end{equation}
\begin{equation}
S_{LR}^{-+,+-}(\omega) = \frac{16 e^2 \alpha_-^2 \alpha_+^2 \Gamma_L \Gamma_R}{2 z_1 + z_2} \frac{z_2^2}{\omega^2 + (2 z_1 + z_2)^2} \; ,
\end{equation}
\begin{equation}
S_{LR}^{+-,-+}(\omega) = \frac{16 e^2 \alpha_-^2 \alpha_+^2 \Gamma_L \Gamma_R}{2 z_1 + z_2} \frac{z_1^2}{\omega^2 + (2 z_1 + z_2)^2} \; .
\end{equation}

For the {\bf large bias voltage} all Andreev states participate and the transfer rates are: $\Gamma_{L +}^{++} = \alpha_+^2 \Gamma_L$, $\Gamma_{R -}^{++} = \alpha_+^2 \Gamma_R$, $\Gamma_{L +}^{--} = \alpha_-^2 \Gamma_L$, $\Gamma_{R -}^{--} = \alpha_-^2 \Gamma_R$, $\Gamma_{L +}^{+-} = \alpha_-^2 \Gamma_L$, $\Gamma_{R -}^{+-} = \alpha_-^2 \Gamma_R$, $\Gamma_{L +}^{-+} = \alpha_+^2 \Gamma_L$, $\Gamma_{R -}^{-+} = \alpha_+^2 \Gamma_R$ and the others are equal $0$. The currents flowing from the L lead through all Andreev bound states are given by
\begin{equation}
J_L^{+-} = \frac{2 e \alpha_-^2 \Gamma_L z_1 z_2}{\Gamma_N^2} \; ,
\end{equation}
\begin{equation}
J_L^{++} = \frac{2 e \alpha_+^2 \Gamma_L z_1 z_2}{\Gamma_N^2} \; ,
\end{equation}
\begin{equation}
J_L^{-+} = \frac{2 e \alpha_+^2 \Gamma_L z_2^2}{\Gamma_N^2} \; ,
\end{equation}
\begin{equation}
J_L^{--} = \frac{2 e \alpha_-^2 \Gamma_L z_1^2}{\Gamma_N^2} \; .
\end{equation}
The asymmetry between the L and the R normal electrode causes current flowing to/from S lead
\begin{equation}
J_S^{-+} = J_S^{+-} = \frac{2 e \alpha_-^2 \alpha_+^2 z_2 \Delta_\Gamma}{\Gamma_N} \; , \;
\end{equation}
\begin{equation}
J_S^{--} = J_S^{++} = \frac{2 e \alpha_-^2 \alpha_+^2 z_1 \Delta_\Gamma}{\Gamma_N} \; , \;
\end{equation}
\begin{equation}
J_S = 4 e \alpha_-^2 \alpha_+^2 \Delta_\Gamma \; ,
\end{equation}
where $\Delta_\Gamma = \Gamma_L - \Gamma_R$ and $\Gamma_N=\Gamma_L+\Gamma_R$.

For the autocorrelation current function $S_{LL}(\omega)$ its intra-channel components can be expressed as
\begin{align}
& S_{LL}^{+-,+-}(\omega) = \frac{4 e^2 \alpha_-^2 \Gamma_L z_1 z_2}{\Gamma_N^2} \nonumber \\
& - \frac{16 e^2 \alpha_-^4 \Gamma_L^2}{\Gamma_N^3} \left( \frac{z_1^2 z_2 \Delta_\alpha \Delta_\Gamma}{\omega^2 + \Gamma_N^2} + \frac{2 z_1^3 z_2}{\omega^2 + 4 \Gamma_N^2} \right) \; ,
\end{align}
\begin{align}
& S_{LL}^{++,++}(\omega) = \frac{4 e^2 \alpha_+^2 \Gamma_L z_1 z_2}{\Gamma_N^2} \nonumber \\
& + \frac{16 e^2 \alpha_+^4 \Gamma_L^2}{\Gamma_N^3} \left( \frac{z_1 z_2^2 \Delta_\alpha \Delta_\Gamma}{\omega^2 + \Gamma_N^2} - \frac{2 z_1 z_2^3}{\omega^2 + 4 \Gamma_N^2} \right) \; ,
\end{align}
\begin{align}
& S_{LL}^{-+,-+}(\omega) = \frac{4 e^2 \alpha_+^2 \Gamma_L z_2^2}{\Gamma_N^2} \nonumber \\
& - \frac{16 e^2 \alpha_+^4 \Gamma_L^2}{\Gamma_N^3} \left(\frac{z_2^3 \Delta_\alpha \Delta_\Gamma}{ \omega^2 + \Gamma_N^2 } + \frac{2 z_1 z_2^3}{ \omega^2 + 4 \Gamma_N^2 }\right) \; ,
\end{align}
\begin{align}
& S_{LL}^{--,--}(\omega) = \frac{4 e^2 \alpha_-^2 \Gamma_L z_1^2}{\Gamma_N^2} \nonumber \\
& + \frac{16 e^2 \alpha_-^4 \Gamma_L^2}{\Gamma_N^3} \left( \frac{z_1^3 \Delta_\alpha \Delta_\Gamma}{ \omega^2 + \Gamma_N^2 } - \frac{2 z_1^3 z_2}{ \omega^2 + 4 \Gamma_N^2 }\right) \; .
\end{align}
and the inter-channel components
\begin{align}
S_{LL}^{-+,+-}(\omega) = \frac{4 e^2 \alpha_-^2 \alpha_+^2 \Gamma_L^2}{\Gamma_N^3} & \left[ \frac{z_2^2 ( 4 z_1^2 + \Delta_\alpha^2 \Delta_\Gamma^2 )}{\omega^2 + \Gamma_N^2} \right. \nonumber \\
& + \left. \frac{4 z_1 z_2 ( z_1^2 + z_2^2 )}{\omega^2 + 4 \Gamma_N^2} \right] \; ,
\end{align}
\begin{align}
S_{LL}^{++,--}(\omega) = \frac{4 e^2 \alpha_-^2 \alpha_+^2 \Gamma_L^2}{\Gamma_N^3} & \left[ \frac{z_1^2 ( 4 z_2^2 + \Delta_\alpha^2 \Delta_\Gamma^2 )}{\omega^2 + \Gamma_N^2} \right. \nonumber \\
& + \left. \frac{4 z_1 z_2 ( z_1^2 + z_2^2 )}{\omega^2 + 4 \Gamma_N^2} \right] \; ,
\end{align}
\begin{align}
S_{LL}^{-+,++}(\omega) = \frac{4 e^2 \alpha_+^4 \Gamma_L^2}{\Gamma_N^3} & \left[ \frac{z_2^2 ( - 4 z_1 z_2 + \Delta_\alpha^2 \Delta_\Gamma^2 )}{\omega^2 + \Gamma_N^2} \right. \nonumber \\
& + \left. \frac{8 z_1 z_2^3}{\omega^2 + 4 \Gamma_N^2} \right] \; ,
\end{align}
\begin{align}
S_{LL}^{+-,--}(\omega) = \frac{4 e^2 \alpha_-^4 \Gamma_L^2}{\Gamma_N^3} & \left[ \frac{z_1^2 ( - 4 z_1 z_2 + \Delta_\alpha^2 \Delta_\Gamma^2 )}{\omega^2 + \Gamma_N^2} \right. \nonumber \\
& + \left. \frac{8 z_1^3 z_2}{\omega^2 + 4 \Gamma_N^2} \right] \; ,
\end{align}
\begin{align}
& S_{LL}^{-+,--}(\omega) = S_{LL}^{+-,++}(\omega) = \nonumber \\
& \frac{8 e^2 \alpha_-^2 \alpha_+^2 \Gamma_L^2}{\Gamma_N^3} \left[ \frac{z_1 z_2 \Delta_\alpha^2 \Delta_\Gamma^2}{\omega^2 + \Gamma_N^2} - \frac{2 z_1 z_2 ( z_1^2 + z_2^2 )}{\omega^2 + 4 \Gamma_N^2} \right] \; ,
\end{align}
where $\Delta_\alpha = \alpha_-^2 - \alpha_+^2$. Notice that one can easily obtain formulae for the currents and their correlation functions for the R junction by changing $L \rightarrow R$ and $\pm \rightarrow \mp$ in the above equations.

The cross correlation function $S_{LR}(\omega)$ is given by
\begin{equation}
S_{LR}(\omega) = -\frac{4 e^2 \Gamma_L \Gamma_R}{\Gamma_N} \frac{( z_1^2 + z_2^2 ) \Delta_\alpha^2}{\omega^2 + \Gamma_N^2} \; ,
\end{equation}
and its components are
\begin{align}
S_{LR}^{+-,+-}(\omega) = \frac{4 e^2 \alpha_-^4 \Gamma_L \Gamma_R}{\Gamma_N^3} & \left[ - \frac{z_2^2 ( 4 z_1^2 + \Delta_\alpha^2 \Delta_\Gamma^2 )}{\omega^2 + \Gamma_N^2} \right. \nonumber \\
& - \left. \frac{4 z_1 z_2 ( z_1^2 + z_2^2 )}{\omega^2 + 4 \Gamma_N^2} \right] \; ,
\end{align}
\begin{align}
S_{LR}^{-+,-+}(\omega) = \frac{4 e^2 \alpha_+^4 \Gamma_L \Gamma_R}{\Gamma_N^3} & \left[ - \frac{z_2^2 ( 4 z_1^2 + \Delta_\alpha^2 \Delta_\Gamma^2 )}{\omega^2 + \Gamma_N^2} \right. \nonumber \\
& - \left. \frac{4 z_1 z_2 ( z_1^2 + z_2^2 )}{\omega^2 + 4 \Gamma_N^2} \right] \; ,
\end{align}
\begin{align}
S_{LR}^{++,++}(\omega) = \frac{4 e^2 \alpha_+^4 \Gamma_L \Gamma_R}{\Gamma_N^3} & \left[ - \frac{z_1^2 ( 4 z_2^2 + \Delta_\alpha^2 \Delta_\Gamma^2)}{\omega^2 + \Gamma_N^2} \right. \nonumber \\
& - \left. \frac{4 z_1 z_2 ( z_1^2 + z_2^2 )}{\omega^2 + 4 \Gamma_N^2} \right] \; ,
\end{align}
\begin{align}
S_{LR}^{--,--}(\omega) = \frac{4 e^2 \alpha_-^4 \Gamma_L \Gamma_R}{\Gamma_N^3} & \left[ - \frac{z_1^2 ( 4 z_2^2 + \Delta_\alpha^2 \Delta_\Gamma^2)}{\omega^2 + \Gamma_N^2} \right. \nonumber \\
& - \left. \frac{4 z_1 z_2 ( z_1^2 + z_2^2 )}{\omega^2 + 4 \Gamma_N^2} \right] \; ,
\end{align}
\begin{align}
& S_{LR}^{-+,+-}(\omega) = \nonumber \\
& \frac{16 e^2 \alpha_-^2 \alpha_+^2 \Gamma_L \Gamma_R}{\Gamma_N^3} \left( \frac{z_2^3 \Delta_\alpha \Delta_\Gamma}{\omega^2 + \Gamma_N^2} + \frac{2 z_1 z_2^3}{\omega^2 + 4 \Gamma_N^2} \right) \; ,
\end{align}
\begin{align}
& S_{LR}^{+-,-+}(\omega) = \nonumber \\
& \frac{16 e^2 \alpha_-^2 \alpha_+^2 \Gamma_L \Gamma_R}{\Gamma_N^3} \left( \frac{z_1^2 z_2 \Delta_\alpha \Delta_\Gamma}{\omega^2 + \Gamma_N^2} + \frac{2 z_1^3 z_2}{\omega^2 + 4 \Gamma_N^2} \right) \; ,
\end{align}
\begin{align}
& S_{LR}^{--,++}(\omega) = \nonumber \\
& \frac{16 e^2 \alpha_-^2 \alpha_+^2 \Gamma_L \Gamma_R}{\Gamma_N^3} \left( - \frac{z_1^3 \Delta_\alpha \Delta_\Gamma}{\omega^2 + \Gamma_N^2} + \frac{2 z_1^3 z_2}{\omega^2 + 4 \Gamma_N^2} \right) \; ,
\end{align}
\begin{align}
& S_{LR}^{++,--}(\omega) = \nonumber \\
& \frac{16 e^2 \alpha_-^2 \alpha_+^2 \Gamma_L \Gamma_R}{\Gamma_N^3} \left( - \frac{z_1 z_2^2 \Delta_\alpha \Delta_\Gamma}{\omega^2 + \Gamma_N^2} + \frac{2 z_1 z_2^3}{\omega^2 + 4 \Gamma_N^2} \right) \; ,
\end{align}
\begin{align}
& S_{LR}^{++,-+}(\omega) = S_{LR}^{-+,++}(\omega) = \nonumber \\
& \frac{8 e^2 \alpha_+^4 \Gamma_L \Gamma_R}{\Gamma_N^3} \left[ - \frac{z_1 z_2 \Delta_\alpha^2 \Delta_\Gamma^2}{\omega^2 + \Gamma_N^2} + \frac{2 z_1 z_2 ( z_1^2 + z_2^2 )}{\omega^2 + 4 \Gamma_N^2} \right] \; ,
\end{align}
\begin{align}
& S_{LR}^{--,+-}(\omega) = S_{LR}^{+-,--}(\omega) = \nonumber \\
& \frac{8 e^2 \alpha_-^4 \Gamma_L \Gamma_R}{\Gamma_N^3} \left[ - \frac{z_1 z_2 \Delta_\alpha^2 \Delta_\Gamma^2}{\omega^2 + \Gamma_N^2} + \frac{2 z_1 z_2 ( z_1^2 + z_2^2 )}{\omega^2 + 4 \Gamma_N^2} \right] \; ,
\end{align}
\begin{align}
& S_{LR}^{++,+-}(\omega) = S_{LR}^{-+,--}(\omega) = \frac{4 e^2 \alpha_-^2 \alpha_+^2 \Gamma_L \Gamma_R}{\Gamma_N^3} \nonumber \\
\times & \left[ - \frac{z_2^2 ( - 4 z_1 z_2 + \Delta_\alpha^2 \Delta_\Gamma^2 )}{\omega^2 + \Gamma_N^2} - \frac{8 z_1 z_2^3}{\omega^2 + 4 \Gamma_N^2} \right] \; ,
\end{align}
\begin{align}
& S_{LR}^{--,-+}(\omega) = S_{LR}^{+-,++}(\omega) = \frac{4 e^2 \alpha_-^2 \alpha_+^2 \Gamma_L \Gamma_R}{\Gamma_N^3} \nonumber \\
\times & \left[ - \frac{z_1^2 ( - 4 z_1 z_2 + \Delta_\alpha^2 \Delta_\Gamma^2 )}{\omega^2 + \Gamma_N^2} - \frac{8 z_1^3 z_2}{\omega^2 + 4 \Gamma_N^2} \right] \; .
\end{align}

\begin{acknowledgments}
This research was financed by National Science Centre (Poland) under the project numbers 2016/21/B/ST3/02160 (BRB, GM) and 2017/27/B/ST3/01911 (TD, KIW).
\end{acknowledgments}